\documentclass[showpacs,preprintnumbers,twocolumn,aps,prl,reprint,superscriptaddress,nobibnotes,nofootinbib]{revtex4-1}

\usepackage{amsmath,amssymb,hyperref,times,graphicx,bm,bbold,mathrsfs}
\usepackage{bibunits}
\usepackage{subfig}
\usepackage[export]{adjustbox}
 \usepackage{enumitem}
\DeclareMathOperator{\Tr}{Tr}
\usepackage{physics}

\usepackage{xcolor}

\newcommand{\be}{\begin{equation}}
\newcommand{\ee}{\end{equation}}
\newcommand{\bea}{\begin{eqnarray}}
\newcommand{\eea}{\end{eqnarray}}

\def\up{\uparrow}
\def\dn{\downarrow}
\def\<{\langle}
\def\>{\rangle}

\providecommand{\up}{\uparrow}
\providecommand{\dn}{\downarrow}

\newcommand{\para}[1]{\left(#1\right)}

\newcommand{\COMMENT}[1]{}

\begin{document}

\begin{bibunit}

\title{An Amelioration for the Sign Problem: Adiabatic Quantum Monte Carlo}

\author{\firstname{Mohammad-Sadegh} \surname{Vaezi}}
\affiliation{\mbox{Pasargad Institute for Advanced Innovative Solutions (PIAIS) , Tehran 19916-33361, Iran}}
\author{\firstname{Amir-Reza} \surname{Negari}}
\affiliation{\mbox{Department of Physics, Sharif University of Technology, Tehran 14588-89694, Iran}}
\author{\firstname{Amin} \surname{Moharramipour}}
\affiliation{\mbox{Department of Physics, Sharif University of Technology, Tehran 14588-89694, Iran}}
\author{\firstname{Abolhassan} \surname{Vaezi}}
\email{Corresponding author. Email address: vaezi@sharif.edu}
\affiliation{\mbox{Department of Physics, Sharif University of Technology, Tehran 14588-89694, Iran}}

\begin{abstract}
We introduce the adiabatic quantum Monte Carlo (AQMC) method, where we gradually crank up the interaction strength, as an amelioration of the sign problem. It is motivated by the adiabatic theorem and will approach the true ground-state if the evolution time is long enough. We demonstrate that the AQMC enhances the average sign exponentially such that low enough temperatures can be accessed and ground-state properties probed. It is a controlled approximation that satisfies the variational theorem and provides an upper bound for the ground-state energy. We first benchmark the AQMC vis-à-vis the undoped Hubbard model on the square lattice which is known to be sign-problem-free within the conventional quantum Monte Carlo formalism. Next, we test the AQMC against the density-matrix-renormalization-group approach  for the doped four-leg ladder Hubbard model and demonstrate its remarkable accuracy. As a nontrivial example, we apply our method to the Hubbard model at $p=1/8$ doping for a $16\times 8$ system and discuss its ground-state properties. We finally utilize our method and demonstrate the emergence of $U(1)_2\sim SU(2)_1$ topological order in a strongly correlated Chern insulator.
\end{abstract}

\maketitle

\noindent {\bf Introduction.}---
Quantum Monte Carlo (QMC) is one of the most powerful tools in computational physics~\cite{blankenbecler1981monte,White_QMC_1989a,von1992quantum,leinweber2000quantum,foulkes2001quantum,gezerlis2013quantum}. It maps an interacting problem into an ensemble over infinitely many non-interacting problems through the Hubbard-Stratonovich (HS) transformations~\cite{hirsch1983discrete}. Every space-time realization of the HS fields defines a path integral which can be evaluated exactly. 
For fermionic models, it usually amounts to computing a determinant. 
The Metropolis-Hastings algorithm is then employed to sample the HS fields according to their weights, namely the corresponding path integrals~\cite{assaad2002quantum,santos2003introduction}. These weights are in general not positive definite and their signs can fluctuate strongly. It has been shown that for a system with volume $V$ and at temperature $T$: $\left<\rm sign\right> \propto \exp\para{-f V/T}$, where $f$ is a model-dependent constant with units of free energy density~\cite{troyer2005computational}. 
Furthermore, the number of samplings required to achieve a desired accuracy scales with the average sign (of the weights) 
as $1/{\langle \rm sign\rangle^{2}}$. Therefore, except for a restricted class of models which are guaranteed to have positive weights~\cite{berg2012sign,li2018numerical,berg2019monte,varney2009quantum,assaad2013pinning,paiva2004critical,li2015solving,li2019sign,li2016majorana,wang2015split,wei2016majorana,huffman2017fermion,alet2016sign}, the sign problem limits the applicability of the QMC approach to high temperatures and small systems~\cite{loh1990sign,troyer2005computational}. 

Since the sign problem is NP-hard~\cite{troyer2005computational} and does not have a general solution, it can be at best alleviated~\cite{umrigar2007alleviation,vaezi2018unified,wan2020mitigating,hangleiter2020easing,levy2019mitigating}. In this letter, we introduce the adiabatic quantum Monte Carlo (AQMC) algorithm as a novel tool to mitigate the sign problem and boost the average sign exponentially. In this formalism, we start with a trial density matrix and evolve it using a time-dependent Hamiltonian with a gradually increasing interaction strength. 
The adiabatic theorem guarantees that the evolved density matrix will approach the true ground-state provided the evolution time is long enough~\cite{nenciu1980adiabatic,wu2005validity,avron1999adiabatic,liu2013quasi}. 
The main observation behind our method is that the average interaction strength is lowered in this method compared to the conventional QMC. It is empirically known that 
the aforementioned constant $f$ is linearly proportional to the (average) interaction strength. Accordingly, the average sign will improve exponentially in the AQMC approach. 

The idea of adiabaticity can be applied to all different types of QMC,  e.g., determinant QMC (DQMC)~\cite{santos2003introduction,assaad2002quantum}, continuous time QMC~\cite{rubtsov2005continuous,gull2011continuous}, constrained-path QMC~\cite{zhang1997constrained}, stochastic series expansions~\cite{sandvik1999stochastic}, etc. Furthermore, it is extremely straightforward to implement the AQMC by minimally modifying the available QMC codes. In the supplemental material (SM), we have provided a simple implementation of the AQMC algorithm in MATLAB which can be used to verify or extend our results.

$~$

\noindent {\bf Algorithm.}---
To elaborate on the details of the AQMC, let us study the nearest neighbor Hubbard model on the square lattice with the following Hamiltonian:
\bea
H_U = &&-t_1 \sum_{\langle{\bf ij}\rangle ,\sigma} \hat{c}_{\bf i,\sigma}^\dag \hat{c}_{\bf j,\sigma} -\mu\sum_{{\bf i},\sigma} \hat{n}_{\bf i,\sigma}\cr
 &&+U\sum_{{\bf i}}\para{\hat{n}_{\bf i,\up}-\frac{1}{2}}\para{\hat{n}_{\bf i,\dn}-\frac{1}{2}},~~~\label{Eq1}
\eea
where $\langle{\bf ij}\rangle$ denotes the nearest neighbor sites, $\hat{c}_{\bf i,\sigma}$ the annihilation operator of a spin $\sigma$ electron at site $i$, and $\hat{n}_{\bf i,\sigma} = \hat{c}_{\bf i,\sigma}^\dag \hat{c}_{\bf i,\sigma}$ is the associated number operator. Throughout this letter, we choose $t_1=1$. The above model is proven to be sign-problem-free at half filling ($\mu = 0$) on bipartite lattices due to its time reversal (TR) and particle-hole (PH) symmetries~\cite{wu2005sufficient}. However, finite doping $\abs{\mu}>0$ breaks the PH symmetry and introduces sign problem~\cite{iglovikov2015geometry}. Motivated by the adiabatic theorem, we consider the following time-dependent auxiliary Hamiltonian: 
\bea
 H_{\mathscr{U}\para{\tau}} := &&-t_1 \sum_{\langle{\bf ij}\rangle ,\sigma} \hat{c}_{\bf i,\sigma}^\dag \hat{c}_{\bf j,\sigma} -\mu'\sum_{{\bf i},\sigma} \hat{n}_{\bf i,\sigma}\cr
 &&+\mathscr{U}\para{\tau}\sum_{{\bf i}}\para{\hat{n}_{\bf i,\up}-\frac{1}{2}}\para{\hat{n}_{\bf i,\dn}-\frac{1}{2}},~~~\label{Eq2}
 \eea
and a trial (initial) wave-function $\left|\Psi_{T}\right\rangle$ that has a non-vanishing overlap with the true ground-state of $H_{U}$, $\left| g\right\rangle_U$. The adiabatic theorem states that the following relation holds provided $H_{\mathscr{U}\para{\tau}}$ has a finite gap throughout the evolution (which is always true for finite discrete systems), $\mathscr{U}\para{\tau}$ is a slowly varying function and $\lim_{\tau \to \beta/2}\mathscr{U}\para{\tau}= U$:
\bea
\left| g\right\rangle_U =  \lim_{\beta \to \infty} \mathcal{T}_{\tau}e^{-\int^{\beta/2}_0  H_{\mathscr{U}\para{\tau}}d\tau} ~ \left|\Psi_{T}\right\rangle\label{Eq3}
\eea
where, $\mathcal{T}_{\tau}$ stands for the (imaginary) time ordering. Since the total electron number commutes with $H_{\mathscr{U}}$ independent of $\mathscr{U}$, $\mu'$ (which enforces the electron density and is different from $\mu$) can be time-independent.  It is straightforward to reformulate the above relation using the density matrix formalism. Defining $\rho_U = \left| g\right\rangle_U \left\langle g\right|_U$, we have:
\bea
&&\rho_U =  \lim_{\beta \to \infty} \mathcal{T}_{\tau}e^{-\int^{\beta/2}_0  H_{\mathscr{U}\para{\tau}}d\tau} ~ \rho_{T}~ {\mathcal{T}_{\tau}}e^{-\int_{0}^{\beta/2}  H_{\overline{\mathscr{U}}\para{\tau}}d\tau},~~~\label{Eq4}
\eea
where, $\overline{\mathscr{U}}\para{\tau}  = \mathscr{U}\para{\beta/2-\tau}$, and $\rho_{T}$ is a trial density matrix (where at least one of its nontrivial eigenstates has a non-vanishing overlap with $\left| g\right\rangle_U$). Obviously, a valid choice is $\rho_{T} =\mathbb{1}$.

The next step is to discretize the (imaginary) time axis and split the evolution time, $\beta/2$ into $N$ steps, which leads to the following approximation of Eq.~\ref{Eq4}:
\bea
\rho_U \approx e^{-\Delta \tau H_{\mathscr{U}_N}}\cdots e^{-\Delta \tau H_{\mathscr{U}_1}}\rho_{T} e^{-\Delta \tau H_{\mathscr{U}_1}}\cdots e^{-\Delta \tau H_{\mathscr{U}_{N}}},~~~\label{Eq5}
\eea
in which $\Delta \tau = \frac{\beta}{2N}$, and $\mathscr{U}_{l} := \mathscr{U}\para{l\Delta \tau}$. After discretization, we employ the second-order Trotter-Suzuki decomposition to rewrite the exponential factors as:
\bea
e^{-\Delta \tau \para{H_{K} + H_{I,l}}} = e^{- \frac{\Delta \tau }{2}H_{K}} e^{-\Delta \tau H_{I,l}} e^{- \frac{\Delta \tau }{2}H_{K}} + O(\Delta \tau^3).~~~~~\label{Eq6}
\eea
where $H_{I,l} = \mathscr{U}_{l}\sum_{{\bf i}}\para{\hat{n}_{\bf i,\up}-\frac{1}{2}}\para{\hat{n}_{\bf i,\dn}-\frac{1}{2}}$, and $H_K$ represents the quadratic part of Eqs.~\ref{Eq1} and \ref{Eq2}. We then apply the Hirsch-Hubbard-Stratonovich transformation~\cite{hirsch1983discrete} to the Hubbard interaction after which: 
\bea
e^{-\Delta \tau \mathscr{U}_l \para{\hat{n}_{\bf i,\up}-\frac{1}{2}}\para{\hat{n}_{\bf i,\dn}-\frac{1}{2}}} = \frac{e^{-\frac{\Delta \tau}{4}\mathscr{U}_l}}{2}\sum_{\sigma = \pm} e^{\sigma \lambda_l  \para{\hat{n}_{\bf i,\up}-\hat{n}_{\bf i,\dn}}}, ~~\label{Eq7}
\eea
where $\cosh\para{\lambda_l} = e^{\Delta \tau \mathscr{U}_l /2}$. The remaining steps are exactly identical to the regular DQMC~\cite{SM}.

In this letter, we make two different choices for the trial density matrix: (I) $\rho_T = \mathbb{1}$, and (II) $\rho_T = e^{-\beta_T H_K}$, i.e., a free fermion thermal ensemble at temperature $1/\beta_T$.
Furthermore, we consider a bounded linear time dependence for the instantaneous onsite couplings: $\mathscr{U}\para{\tau} = {\rm min}\para{U,\mathscr{U}_{\rm max} \frac{2\tau}{\beta}}$. The convergence criteria for the AQMC is the flattening  and convergence of energy and other desired observables against increasing $\beta$ further.

Unlike a number of other powerful tools designed to mitigate the sign problem, e.g., constrained-path QMC~\cite{carlson1999issues},  AQMC satisfies the variational theorem and therefore provides an upper bound for the ground-state energy. It is thus a controlled approximation in the sense that increasing the evolution/projection time, or optimizing the parameters of the trial density matrix will increase the accuracy of the method and will bring us closer to the true ground-state while never crossing it.
Moreover, for symmetry preserving trial density matrices, AQMC is an unbiased method and respects the ergodicity of the problem.

\noindent {\bf Benchmarking AQMC}.--- In the SM, we have benchmarked the AQMC algorithm by studying two cases with known exact results. We first compared its performance and accuracy with the regular DQMC for the Hubbard model on the square lattice at half filling, which is sign problem free, for a $16\times 2$ square lattice subject to $U=4$. As Fig. S6 of the SM shows, the AQMC recovers the exact results for sufficiently long evolution times ($\beta$).  To further validate our method, we then considered the doped Hubbard model in the SM which is known to suffer from the sign problem. There, we studied a $16\times 4$ square lattice on a cylinder, for $U=4$ and at $p=1/8$ doping . Fortunately, an accurate estimate of the ground-state energy for this problem is available from the DMRG approach.~\cite{ehlers2017hybrid}. Again, our AQMC approach yields exact results for $\beta \gtrapprox 10$ where the energy (per site) converges to $E= -1.0182 \pm 0.0003$ (see Fig. S7). Besides these two cases, in the SM, we have shown the superiority of our approach over the DQMC for other fillings, $U$ values, lattice geometries and system sizes.

\noindent {\bf Results}.--- In this section, with the help of AQMC, we study two different strongly interacting models: (I) $p=1/8$ doped Hubbard model on a periodic $16\times 8$ square lattice and demonstrate the absence of superconductivity in the pure Hubbard model ($t'=0$). (II) A highly interacting spin degenerate Chern insulator where we prove an emergent topological order identical to that of the bosonic $1/2$ Laughlin state.

\begin{figure}[t]
\centering
 \includegraphics[width=8.5cm]{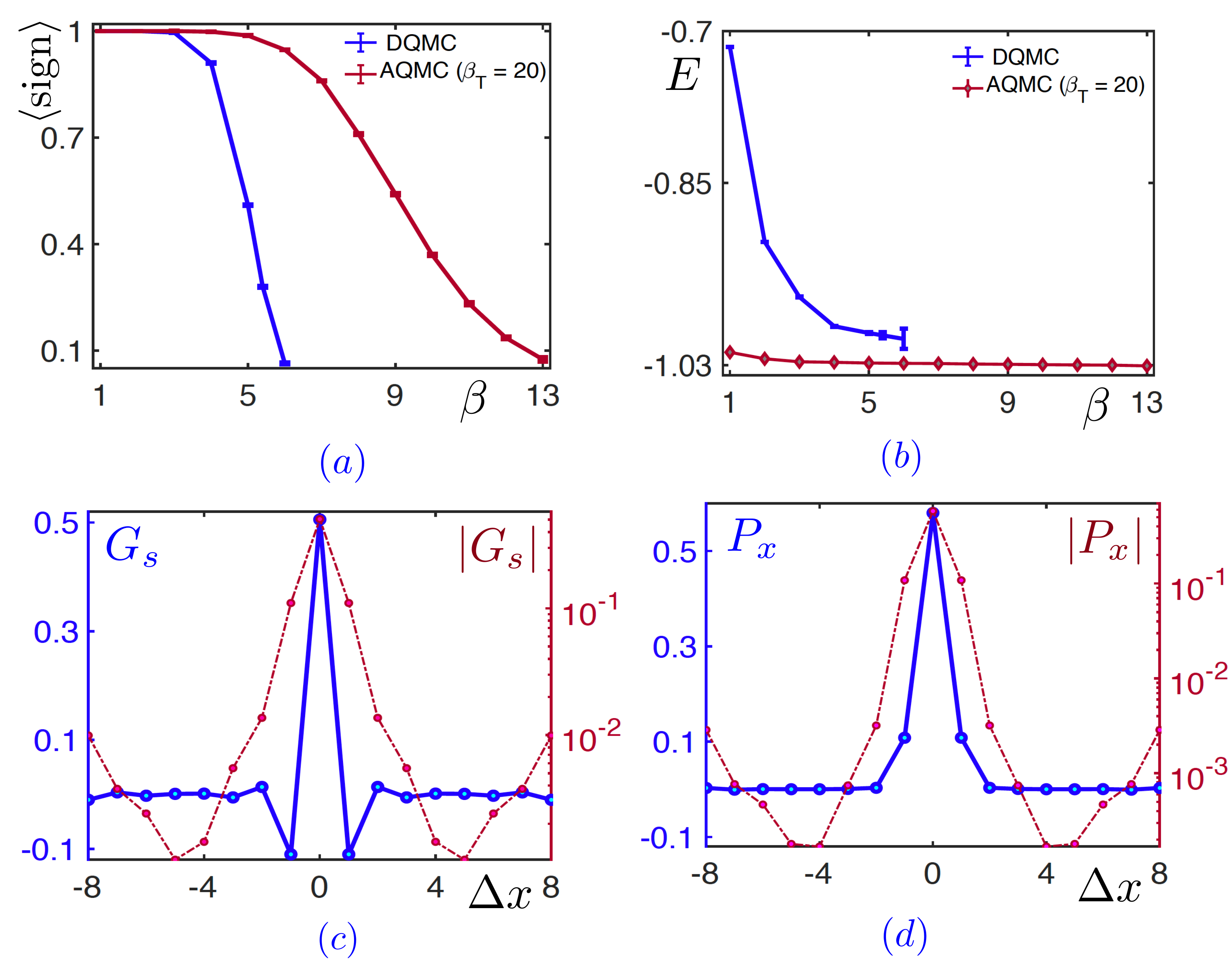}%
\caption{\raggedright (a)-(b) Comparing the average sign and energy of the regular DQMC and those of the AQMC ($\beta_T = 20$) against $\beta$  for a $16\times 8$ periodic system subject to $U=4$ Hubbard interaction at $p=1/8$ doping.
(c)-(d) Magnetic and superconducting correlation functions along the $x$ axis obtained via AQMC with $\beta=12$, and $\beta_T = 20$ for the $16\times 8$ sample at $p=1/8$ and $U=4$. The blue plots are in normal-normal scale and depict Green's functions and the red ones are in normal-log scale and represent the {\em absolute value} of the corresponding Green's functions. The spin-spin correlation function suggests a stripe order. In contrast, the pair-pair correlation function is relatively much smaller at long distances suggesting the absence of long-range superconducting order.
}
\label{fig3}
\end{figure}

\begin{figure}[t]
\centering
\includegraphics[width=8.5cm]{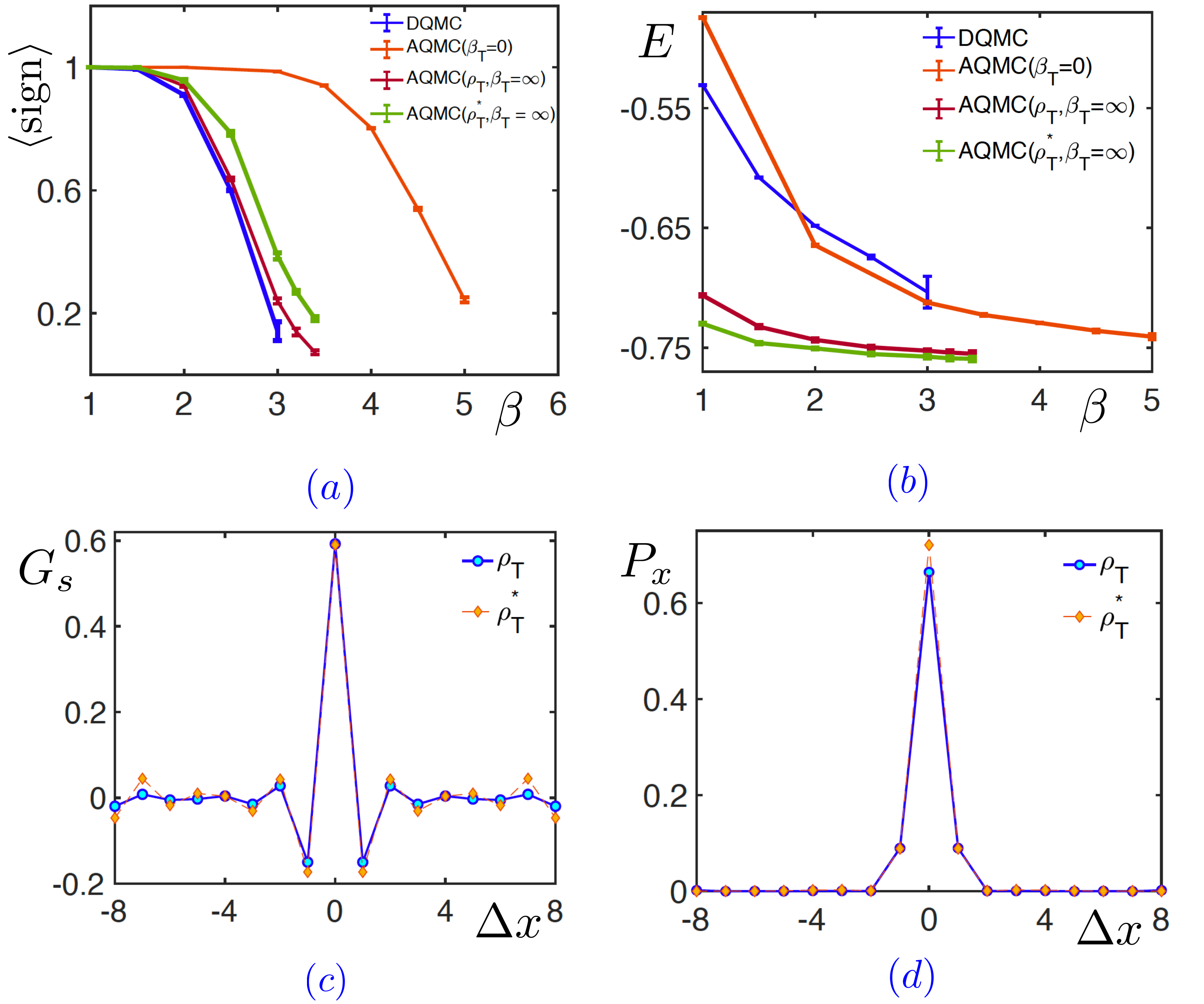}%
\caption{\raggedright AQMC U=8 results for an $16\times 8$ system at $p=1/8$ doping. $\rho_T$ denotes the free Fermi surface trial density matrix, and $\rho_T^*$ the optimized trial state with non-vanishing stripe order parameters. (a)-(b) Comparing the average sign and energy (versus $\beta$) of the AQMC with that of the DQMC methods. The AQMC can reach lower temperatures ($\beta$'s) and yields lower energies.  
(c) The spin-spin and pair-pair correlation functions achieved via AQMC with $\rho_T$ (blue) and $\rho_T^*$ (orange) trial states. In both cases, the pair-pair correlation functions decay considerably faster at long distances, implying a non-superconducting state. Although $\rho_T^*$ corresponds to a slightly lower energy state than $\rho_T$, they both consistently indicate the stripe order and belong to the same phase.}
\label{fig4}
\end{figure}

\noindent {\bf I. $U=4$ and $U= 8$ Hubbard model at $1/8$ doping on a periodic $16\times 8$ system}.---
This filling has been the subject of intense research in the past few decades. There are numerous studies suggesting a plethora of  various competing phases
for the Hubbard model at $p=1/8$ doping level with close energies including the d-wave superconductivity, and stripe order phases~\cite{kivelson2003detect,yang2011proximity,huang2017numerical,dolfi2015pair,ehlers2017hybrid,jiang2019superconductivity,dodaro2017intertwined,darmawan2018stripe,zheng2016ground,zheng2017stripe,leblanc2015solutions,corboz2014competing}.
Our method can conveniently handle the system size and the boundary conditions considered here.  Moreover, due to its enhanced average sign we can achieve the ground-state properties. This can be verified from Fig.~\ref{fig3}b, which indicates that the average energy for $U=4$ nearly plateaus around $E_g = -1.0305 \pm 0.0005$. The regular DQMC is obviously unable to reach beyond $\beta = 5.4$ due to its appalling sign problem (see Fig.~\ref{fig3}a). On the other hand, DMRG is inapplicable for this system size and geometry. The torus boundary conditions aside, $N_y=8$ requires gigantic bond dimensions for the convergence (as large as $10^6$) which in currently unaffordable even on the best available computational facilities.

Our spin-spin correlation function clearly points towards the stripe order formation. As Fig.~\ref{fig3}c shows, we evidence a $\pi$ phase shift in the correlations after $\abs{\Delta x}=4$. This correlation function is consistent with a magnetization of the form $\langle S_z\para{x,y}\rangle = m\cos\para{Q_x x + Q_y y}$, where $\para{Q_x,Q_y} = \para{\frac{7}{8},1}\pi$. We have also plotted the superconducting (pair-pair) correlation function $P_{x}\para{\bf r}=  \langle \Delta_{x}\para{\bf i}^\dag \Delta_{x}\para{\bf i + r}\rangle $ (averaged over all possible $\bf i$'s), where
 $\hat{\Delta}_{x}\para{\bf i} \equiv \hat{c}_{\bf i , \up} \hat{c}_{{\bf i} + \hat{x}, \dn}- \hat{c}_{\bf i , \dn} \hat{c}_{{\bf i} + \hat{x}, \up}$. From the log-normal plots, it is clear that $P_x$ is substantially more suppressed than $G_s$ at long distances. Thus, we conclude that the ground-state is not a superconductor~\cite{SM}. This is consistent with the growing agreement that for the pure Hubbard model ($t'=0$), the stripe order wins the competition and the superconductivity is absent at this filling~\cite{qin2020absence}.
 
We now apply the AQMC to $U=8$. As Fig. \ref{fig4} implies, the AQMC outperforms the regular DQMC and achieves lower energies. 
Moreover, when we assume no ordering in $\rho_T$, and start with a free Fermi surface as our trial (initial) state, the energy versus $\beta$ curve does not converge for $\langle {\rm sign }\rangle > 0.1$. 
The lowest energy we obtain with this initial state is $E=-0.755 \pm 0.002$. For this state, similar to $U=4$, the correlation functions point toward the stripe order phase. 
However, by modifying $\rho_T$ to reflect the observed ordering (which we refer to as $\rho_T^*$), the convergence to the true ground-state can be achieved (with $E= -0.759\pm 0.002$).

\begin{figure}[t]
\centering
\includegraphics[width=8.5cm]{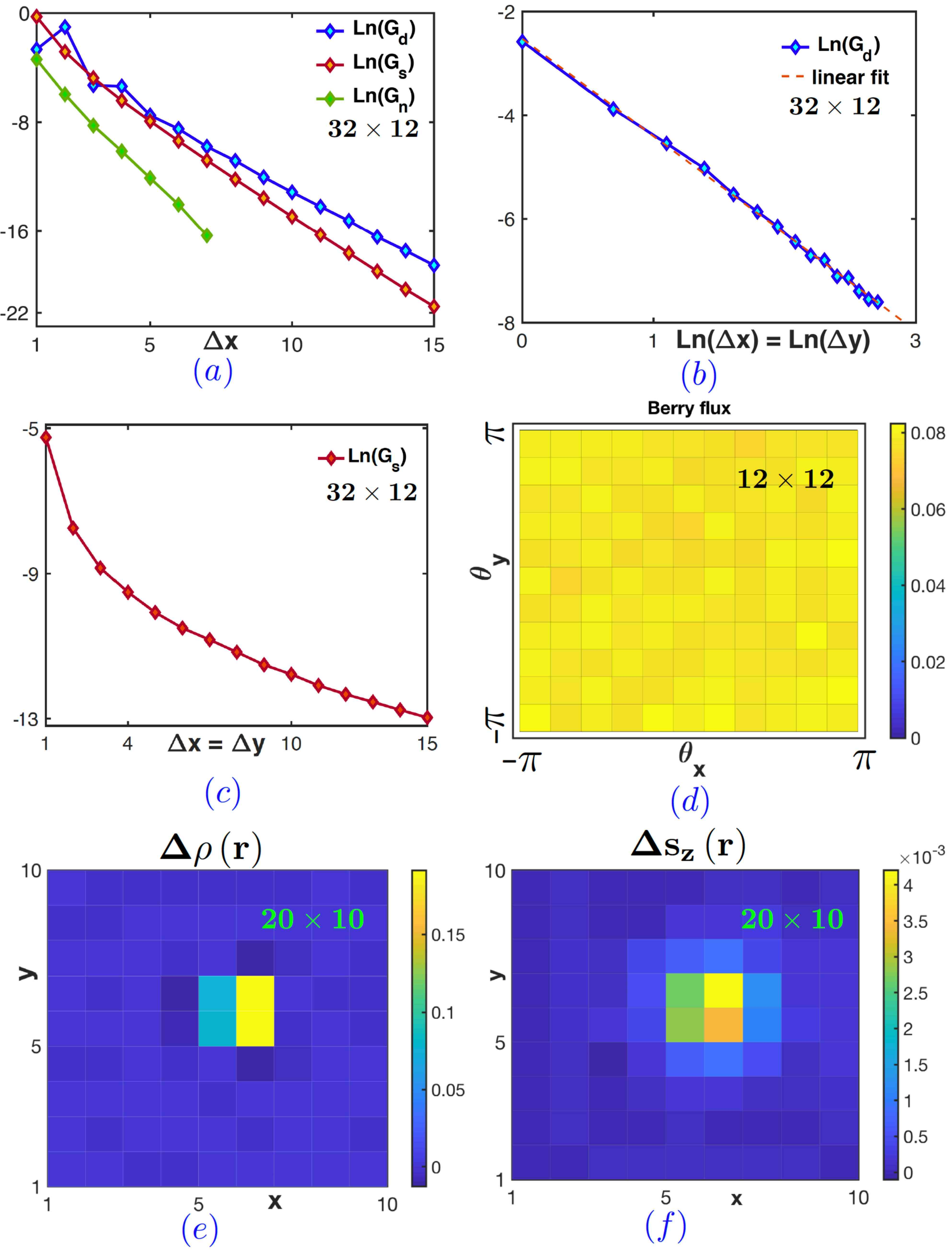}%
\caption{\raggedright The results of AQMC for a correlated spinful Chern insulator ($U=-6$). The system size for each plot is indicated. 
(a) The doublon-doublon, spin-spin, and density-density correlation functions die off exponentially in the bulk. The corresponding decay lengths are all less than a unit cell. (b) The doublon-doublon correlators decay algebraically as $1/r^{2h_d}$ at the edge of the system ($h_d \approx 0.95$), for an edge created along the $\hat{x}+\hat{y}$ direction. We obtain a nearly identical profile for the density-density correlation function consistent with an emergent $SU(2)$ edge symmetry.
(c) Edge spin-spin correlators decay exponentially at long distances (with an enlarged decay length around 4 unit cells). (d) The many body Berry flux is nearly uniform in the entire phase space and its integral, the Berry phase, is fractional and equal to $0.98\pi$ in agreement with the theoretical predictions for a $\nu = 1/2$ Laughline state. (e) For a $\Phi_{\up} = \Phi_{\dn} = hc/2e$, the total accumulated charge (relative to the background charge) equals $0.502$ (in units of $2e$), while in (f) the accumulated spin in negligible ($s_{z,\rm tot} = 0.02$) for $\Phi_{\up} = -\Phi_{\dn} = hc/2e$. Note that only the left half space is shown and fluxes are inserted around the center.}
\label{figCSL}
\end{figure}

\noindent {\bf II. Correlated Chern insulators}.---As our second nontrivial model Hamiltonian, we focus on a spin-degenerate staggered-flux Chern insulator~\cite{neupert2011fractional} with strong onsite Hubbard attraction ($U=-6$) between electrons with opposite spins \cite{SM}. Here, the band-structure for spin up and down electrons contains two nearly flat bands with Chern number $C = \pm 1$ for the valence and conduction bands, respectively.  The system is at half filling, thus the valence band is fully occupied at $U=0$.  This model has a severe sign problem due to the lack of time-reversal symmetry~\cite{wu2005sufficient,SM}. The fate of this model is not fully settled yet~\cite{maciejko2013topological,ruegg2012topological,he2011chiral,hickey2015competing,wu2016quantum,arun2016mean,zhang2017topological} although some evidences in favor of topological order have been found. There are several possible ground-states for this model for strongly interacting case including an s-wave superconductors~\cite{hickey2015competing,zhang2017topological}, a charge/spin density wave~\cite{arun2016mean}, and a nontrivial state with topological order similar to that of $\nu=1/2$ Laughline state~\cite{maciejko2013topological,ruegg2012topological,he2011chiral}. In the latter phase, the fundamental degrees of freedom are charge $q=2e$ doublons ($d_{\bf r} \equiv c_{\bf r, \up} c_{\bf r, \dn}$), thus all spin-carrying operators such as electrons are confined~\cite{SM}. Furthermore, the ground-state is two-fold degenerate on torus geometry, hosts non-trivial chiral edge states described by a $SU(2)_1\equiv U(1)_2$ conformal field theory (CFT), and contains excitations with fractional charge ($q/2$) and fractional statistics ($\pi/2$)~\cite{maciejko2013topological}. Our results below, exclude the former two candidates and supports the topological order. To carefully uncover the nature of the ground-state, we studied systems as large as  $32\times 12$ which only thanks to the AQMC became achievable.

We first exclude the possibility of s-wave superconductivity and CDW order by studying the bulk pair-pair, density-density, and spin-spin correlation functions. As Fig. \ref{figCSL}a suggests, all correlation functions decay exponentially in the bulk implying the absence of any symmetry breaking long range order. Next, we study the previously mentioned two-point correlators at the edge. For $U(1)_2 \equiv SU(2)_1$ CFT, we expect the density-density and pair-pair correlation functions to follow the same profile and both decay algebraically with $h_d = h_n = 1$ conformal dimensions, consistent with our results (see Fig.  \ref{figCSL}b). On the other hand, the topological order requires the spin degrees of freedom to be gapped even at the edge~\cite{SM} and as a result its correlators must decay exponentially everywhere, consistent with our results in Fig. \ref{figCSL}c.
Another non-trivial fingerprint of the $1/2$ Laughline state is its fractional many body Chern number $(C_{\rm MB}=1/2)$~\cite{neupert2011fractional}. The computation of $C_{\rm MB}$ involves applying twisted boundary conditions along $x$ and $y$ directions, and evaluating the overlap between many body wavefunctions with different twist angles~\cite{MoharramipourFQH}. As a result, similar to the single particle calculations, we can define the Berry curvature whose integral over the entire phase space determines the many body Chern number. In Fig. \ref{figCSL}d we plot the Berry flux density which amounts to total Berry phase $\theta_{B} \approx \pi$. Hence $C_{\rm MB} = \frac{\theta_B}{2\pi} \approx 1/2$. 

Finally, theoretical considerations dictate that a quantum flux of doublons, i.e., $\Phi = hc/q=hc/2e$ will trap a nontrivial anyon excitation (known as semions in the $U(1)_2$ phase) with charge $q/2=e$ bound to the flux location. In terms of the underlying microscopic degrees of freedom i.e., electrons, this translates into three different choices: (a) $\Phi_{\up}\para{\bf r} = \Phi_{\dn}\para{\bf r}=hc/2e$, i.e., both electrons couple to the inserted fluxes symmetrically.(b) $\Phi_{\up} = 2\Phi = hc/e$, and $\Phi_{\dn} = 0$. (c) $\Phi_{\dn} = 2\Phi = hc/e$, and $\Phi_{\up} = 0$.  In the $U(1)_2$ topological order spin degrees of freedom are gapped. Consequently, all the above three choices will lead to the same {\em spinless} charge $e$ excitations, while in the non-interacting model, (a) is unacceptable since the fluxes are quantized in unites of $hc/e$, (b) will excite a spin up electron and (c) excites a spin-down electron. Therefore, the fractional excitations of $U(1)_2$ topological state (semions) are quite distinct from spinful electrons. Instead, they are {\bf {\it spinless fractionalized doublons}} with {\bf {\it charge $q/2 = e$}}.
From the above discussion, we can conclude that in $U(1)_2$ topological order: (a) $\Phi_{\up} = \Phi_{\dn}=hc/2e$ excites a semion with charge $q/2=e$. (b) $\Phi_{\up} = -\Phi_{\dn}=hc/2e$ is a trivial {\it spinless} and charge {\it neutral} excitation. Now, let us employ these two diagnostics. In Fig. \ref{figCSL}e, we considered  $\Phi_{\up}\para{\bf r} = \Phi_{\dn}\para{\bf r}=hc/2e$ flux insertions. As we see, the total excess charge is quantized at $q/2=e$ and localized around the inserted flux. More importantly, in Fig. \ref{figCSL}f, we considered $\Phi_{\up}\para{\bf r} = -\Phi_{\dn}\para{\bf r}=hc/2e$ flux insertion. We observe that this corresponds to a neutral excitation with a negligible accumulated spin, consistent with theory.

\noindent {\bf Discussion}.---So far, we considered the simplest form of the AQMC algorithm. Our trial density matrix was simply related to the hopping Hamiltonian as $\rho_{T} = e^{-\beta_T H_K}$. We can in principle consider more general choices for $\rho_{T}$ ($\rho_T^*$), and replace $H_K$ by a variational $H_K^*$ with variational inputs, e.g., order parameters of relevant symmetry breaking phases and optimize the ground-state energy with respect to them. 
Another direction to improve AQMC is to consider an interacting trial density matrix by utilizing the variational QMC algorithm~\cite{vaezi2018unified}. In this approach, we can reach the true ground-state with shorter evolutions. Finally, the profile of $\mathscr{U}\para{\tau}$ can be treated as a variational function. Although we applied the AQMC to the Hubbard model within the DQMC framework, it is not limited to them. We can apply it to other fermionic or bosonic model Hamiltonians or implement it within other QMC frameworks. Among those model Hamiltonians, in particular we would like to mention the multi-layer Hubbard model and the partially flat band systems~\cite{huang2019enhanced,sayyad2020pairing,aoki2020theoretical}. The average sign is already larger in these models and can be boosted further using AQMC such that the ground-state properties will be more accurately captured.

$~$

\noindent {\bf Acknowledgements.}--- We thank Edwin Huang, Sharareh Sayyad, Hongchen Jiang and Ehsan Khatami for helpful discussions and carefully reading our manuscript. AV acknowledges the Gordon and Betty Moore Foundation’s EPiQS Initiative through Grant GBMF4302 and Stanford Center for Topological Quantum Physics for partial financial support and hospitality during the completion of this work. MSV acknowledges the financial support from Pasargad Institute for Advanced Innovative Solutions (PIAIS) under supporting Grant scheme (Project No. SG1-RCM2002-01).

\putbib
\end{bibunit}

\clearpage

\subsection{SUPPLEMENTAL MATERIAL}

In this supplemental material (SM), we first delve into the details of the adiabatic quantum Monte Carlo (AQMC) algorithm. Next, we compare the energy and average sign of the AQMC with those of the standard determinant quantum Monte Carlo (DQMC) for various system sizes, interaction strengths, doping levels, and two different lattice geometries. In section 3 of this SM, we benchmark the AQMC method for two cases with known exact results to prove its accuracy and reliability. In section 4, we present more results for the Hubbard model at $p=1/8$. In section 5, we argue that strong attractions may lead to the emergence of a $\nu=1/2$ Laughline state from correlated Chern insulator. At the end, in section 6, we provide a simple implementation of the determinant and adiabatic quantum Monte Carlo in a single script in the MATLAB environment. 

\section{1. Details of the AQMC algorithms}

Here we discuss the implementation of the AQMC algorithm in more detail. It involves very similar computational steps to a regular determinant quantum Monte Carlo (DQMC).

As it is discussed in the main text, the following expression considered as an approximation to the true density matrix (and becomes exact in the $\beta \to \infty$ limit):

\be \rho_U \approx e^{-\Delta \tau H_{\mathscr{U}_N}}\cdots e^{-\Delta \tau H_{\mathscr{U}_1}}\rho_{T} e^{-\Delta \tau H_{\mathscr{U}_1}}\cdots e^{-\Delta \tau H_{\mathscr{U}_{N}}}.\tag{S1} \ee

The expectation value of any physical operator (e.g., the Green's function) is obtained via the following relation:

\be \ev{G} = \frac{\Tr(G \rho_U)}{\Tr( \rho_U)} . \tag{S2} \ee

After performing the Hubbard-Stratonovich (HS) transformation, it can be shown that the Green's function for spin $\sigma$ electrons can be written as

\be \ev{G^{\sigma}} = \frac{\sum_{\vb{s}}G_{\vb{s}}^{\sigma} \mathcal{W}_{\vb{s}}}{\sum_{\vb{s}} \mathcal{W}_{\vb{s}}} = \frac{\ev{G_{\vb{s}}^{\sigma} \text{sign} (\mathcal{W}_{\vb{s}})}}{\ev{\text{sign}(\mathcal{W}_{\vb{s}})}}, \tag{S3} \ee
where the expectation values over the Hubbard-Stratonovich field can be computed using Metropolis-Hastings sampling algorithm with the acceptance probability of $P(\vb{s} \rightarrow \bar{\vb{s}}) = \min ( \frac{\abs{\mathcal{W_{\bar{\vb{s}}}}}}{\abs{\mathcal{W_{\vb{s}}}}}, 1)$
and $G_{\vb{s}}^{\sigma}$ and $\mathcal{W_{\vb{s}}}$ are given by
\be G_{\vb{s}}^{\sigma} = (I + B_{2N+N'}^{\sigma} \cdots B_{N+1}^{\sigma} B_{N}^{\sigma} \cdots B_{1}^{\sigma})^{-1}, \tag{S4} \label{G_B}\ee

\be \mathcal{W}_{\vb{s}} = (\det G_{\vb{s}}^{\uparrow})^{-1} (\det G_{\vb{s}}^{\downarrow})^{-1}, \tag{S5} \ee
where the $B$ matrices are defined as
\be
B_l^\sigma = e^{-\Delta \tau {\mathsf{K}}} e^{- dig( \lambda_l \sigma s_i(l)) }, \tag{S6}
\ee
with
\be
K_{i j}=
\begin{cases}
   -t & \text{if $i$ and $j$ are nearest neighbors},\\
   -\mu' & \text{if $i$ = $j$}, \\
   0  & \text{otherwise}
\end{cases}
\tag{S7}
,\ee 

\be
\mathscr{U}_l=
\begin{cases}
   \frac{U}{N} (N - l) & \text{if $ l \leq N$},\\
   0 & \text{if $ N < l < N + N' $},\\
   \frac{U}{N} \para{l-\para{N + N'}} & \text{if $ N + N' \leq l$}
\end{cases}
\tag{S8}
\label{Deltaell}
\ee
and $\cosh \lambda_l = e^{\Delta \tau \mathscr{U}_l /2}$. The correlation functions other than the fermionic Green's function can be also found using the Wick's theorem which is valid for each Monte Carlo sampling step.

In order to update Monte Carlo configuration and walk through the Markov chain, we have to check whether the HS spin at each space-time point would be flipped or not.   
To perform this update, we use the Sherman-Morrison update rule which is more efficient than calculating the acceptance probability and the updated Green's function from scratch and only requires $\mathcal{O}(N_s^2)$ operations rather than $\mathcal{O}(N_s^3)$ which is the typical computational complexity of matrix manipulations.

Using Sherman-Morrison formulas, one can prove that the following relations can be applied to find the acceptance ratio and updated Green's function with HS spin at time $l = 1$ and site $i$ flipped.
\be
\alpha^\sigma(i)\equiv e^{-2\lambda_l\sigma 
s_{i}}-1,
\tag{S9}
\ee

\be r^{\sigma}(i) = \frac{\det (\bar{G}^{\sigma})^{-1}}{\det (G^{\sigma})^{-1}} = 1 + \alpha^{\sigma}(i)(1-G^{\sigma}_{ii}), \tag{S10}\ee

\be
\bar{G}_{j k}^\sigma=
G_{j k}^\sigma-
\frac{\alpha^\sigma(i)}{r^{\sigma}(i)} \left[\delta_{j i}-G_{j i}^\sigma\right] G_{i k}^\sigma \tag{S11}
\ee

\begin{figure}[t]
\centering
\includegraphics[width=8 cm]{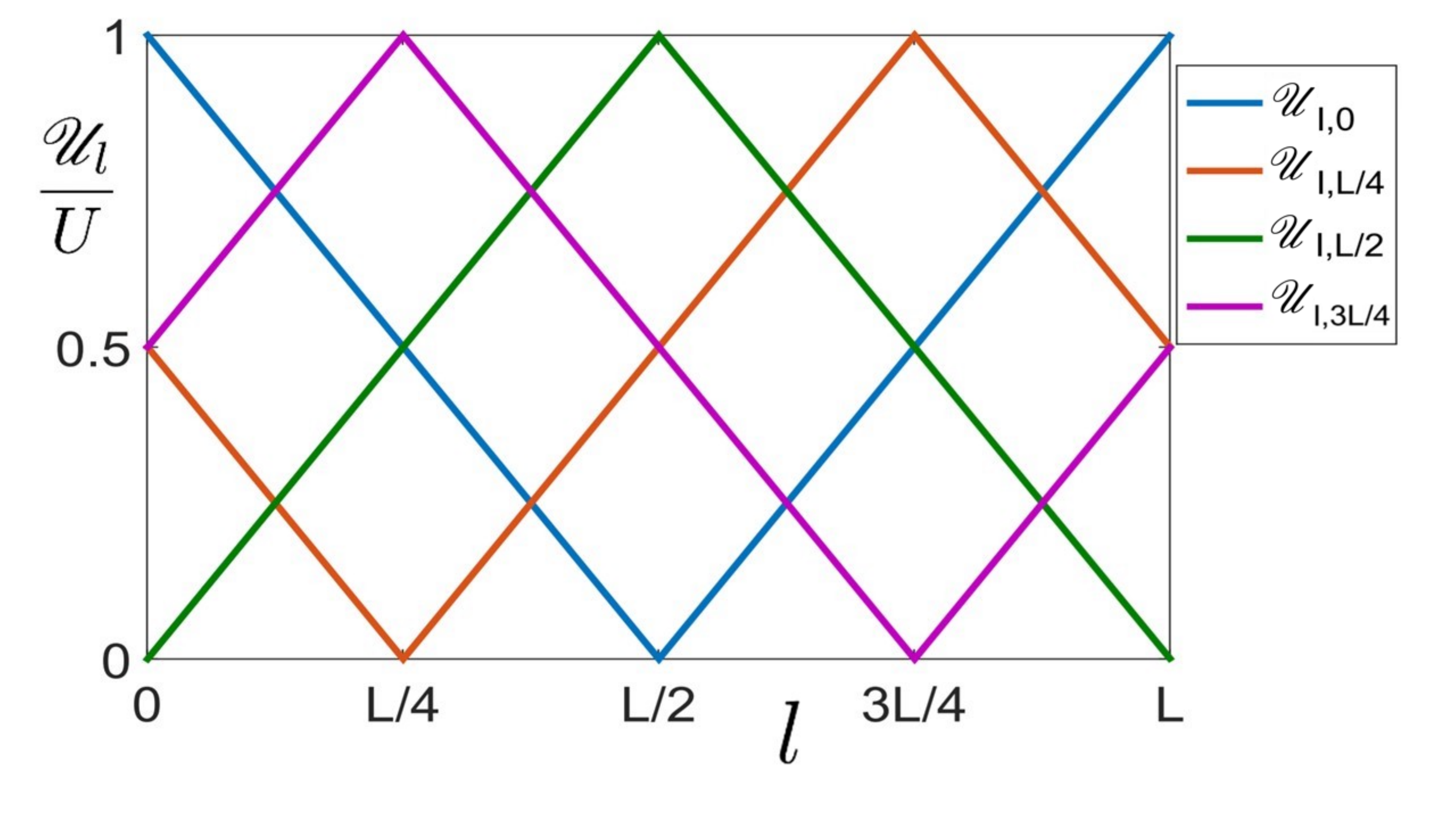}%
\caption{\raggedright Plots of $\mathscr{U}_{l,l_0}$ for some different values of $l_0$ which indicates the number of time slice shifts. For the sake of clarity, here we have considered $\beta_T = 0$ ($N'=0$).}
\label{figS6}
\end{figure}

To update HS spins at time $ l = 2 $, we need to bring the $B_2$ matrix the rightmost place of $B_{2N+N'}^{\vb{\sigma}} \cdots B_{N+1}^{\vb{\sigma}} B_{N}^{\vb{\sigma}} \cdots B_{1}^{\vb{\sigma}}$ 
This can be easily done by the following transformation,
\be
{G}^\sigma(l+1)= B_l^\sigma {G}^\sigma(l) 
[B_l^\sigma]^{-1}
\tag{S12}
\ee
Since the wrap up operation does not change the determinant, Sherman-Morrison updating formulas are still allowed. By repeating this ``update-wrap up" process, one can sweep the imaginary time direction and update all of the HS spins. After each wrap up, the  $\mathscr{U}_l$ gets shifted circularly (see Fig. (\ref{figS6})) and finally after a complete space-time sweep returns to its initial form. Hence, physical measurements have to be done once at the end of each space-time sweep where the $\mathscr{U}_l$ profile is symmetric around $L/2$ ($L = 2N + N'$  is the total number of time steps).

Furthermore, in order to decrease the Trotter error that arises due to time discretization, one could employ second order Trotter-Suzuki decomposition which is  

\be e^{-\Delta \tau K - \Delta \tau V} = e^{-\frac{\Delta \tau K}{2}} e^{-\Delta \tau V} e^{-\frac{\Delta \tau K}{2}} + \mathcal{O}(\Delta \tau ^ 3) \tag{S13} \ee      

and leads to an error of $\mathcal{O}(\Delta \tau^3)$ rather than $\mathcal{O}(\Delta \tau^2)$. Since, $G^\sigma$ is determined in terms of the product of $\mathcal{O}({\frac{1}{\Delta \tau}})$ $B$ matrices, its Trotter error would be $\mathcal{O}(\Delta \tau^2)$ in this case. More specifically, the second order Trotter error is $\mathcal{O} \qty\Big(\frac{\Delta \tau ^3}{24} \qty\big( 2\qty\big[V,[K,V]] + \qty\big[K,[K,V]] ) )$. Since the interaction strength is about several times greater than the hopping amplitude ($t = 1$ in our code), the latter commutator is negligible compared to the former. Therefore, the Trotter error would be proportional to $ \mathscr{U}(\tau)^2 \Delta \tau^3 $ at a single (imaginary) time step. Basically, the overall error is obtained by integrating the errors at single time steps over time. For the AQMC case in which $\mathscr{U}(\tau)$ is a linear function, this error would be proportional to $\mathcal{O}(\beta U^2 \Delta \tau^2 /3)$ which is three times smaller than the error for the regular DQMC ($\mathscr{U}(\tau) = \text{const.})$.

 According to the Eq. (\ref{G_B}) which determines $G^{\sigma}$ in terms of $B$ matrices, it is easy to show that it is sufficient to modify the Green's function by  $ G^{\sigma}_{ \text{2nd order} } = e^{-\frac{\Delta \tau K}{2}} G^{\sigma}_{\text{1st order}} e^{\frac{\Delta \tau K}{2}} = e^{-\frac{\Delta \tau K}{2}} G^{\sigma} e^{\frac{\Delta \tau K}{2}} $ in order to use the second order Trotter-Suzuki decomposition for measuring physical quantities after each space-time sweep.

\section{2. A systematic comparison between AQMC and DQMC}

Below, in Fig. \ref{figS1} we have applied the AQMC algorithm to $U=4,6,8$, and $p =1/8$ for $N_x=N_y=8$ square geometry. As we see, the AQMC ameliorates the sign problem for all three cases, and achieves a lower energy than the standard DQMC. 

In Fig. \ref{figS2}, we have considered the following three different system sizes:  $4\times4$, $8\times8$, and $12\times12$. In every case, we conclude that it can access a $\beta$ larger the maximum $\beta$ available via the standard DQMC, results an exponentially larger average sign in the same $\beta$, and above all yields a lower energy estimate and therefore has a larger overlap with the true ground-state. 

In Fig. \ref{figS3}, we have considered the following three doping levels: $p=0.2,1/8,0.05$ corresponding to $<n>=0.8,0.875,0.95$ fillings, respectively. Again, for all fillings our expectations are fulfilled.
\begin{figure}[t]
\centering
\includegraphics[width=8 cm]{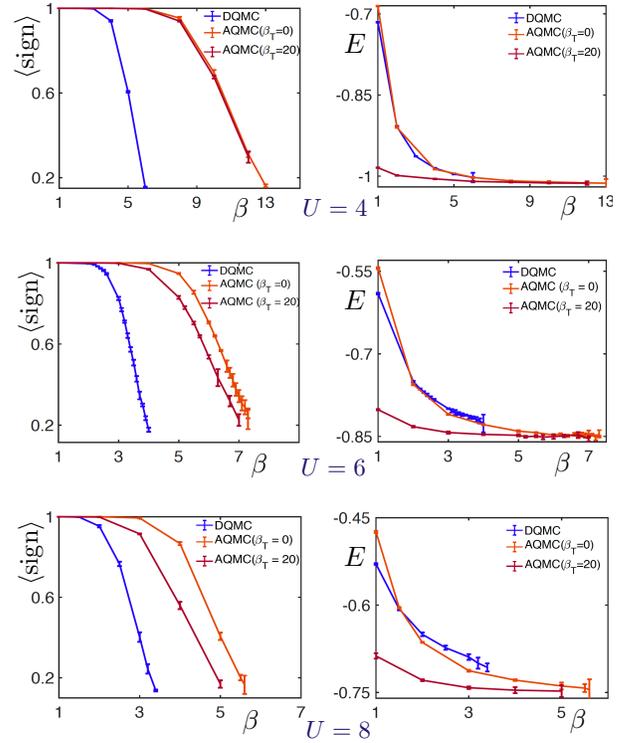}%
\caption{\raggedright Comparing the AQMC with DQMC for $U=4,6,8$ at $p=1/8$ doping ($<n>=0.875$) for an $8\times8$ square geometry.}
\label{figS1}
\end{figure}

\begin{figure}[t]
\centering
\includegraphics[width=8 cm]{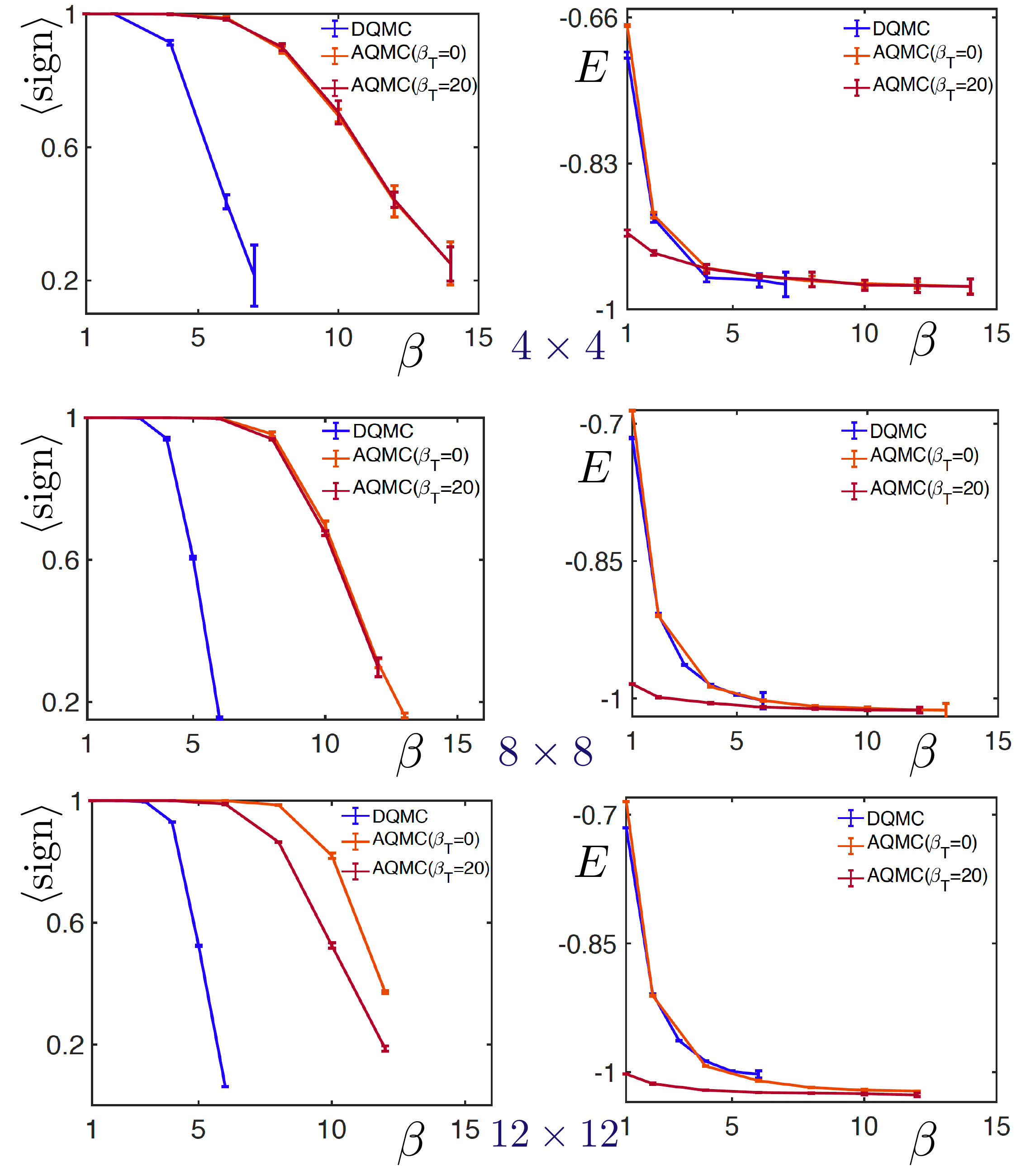}%
\caption{\raggedright Comparing the AQMC with DQMC for $U=4$ at $p=1/8$ doping for $4\times4$, $8\times8$, and $12\times12$ size.}
\label{figS2}
\end{figure}

\begin{figure}[t]
\centering
\includegraphics[width=8 cm]{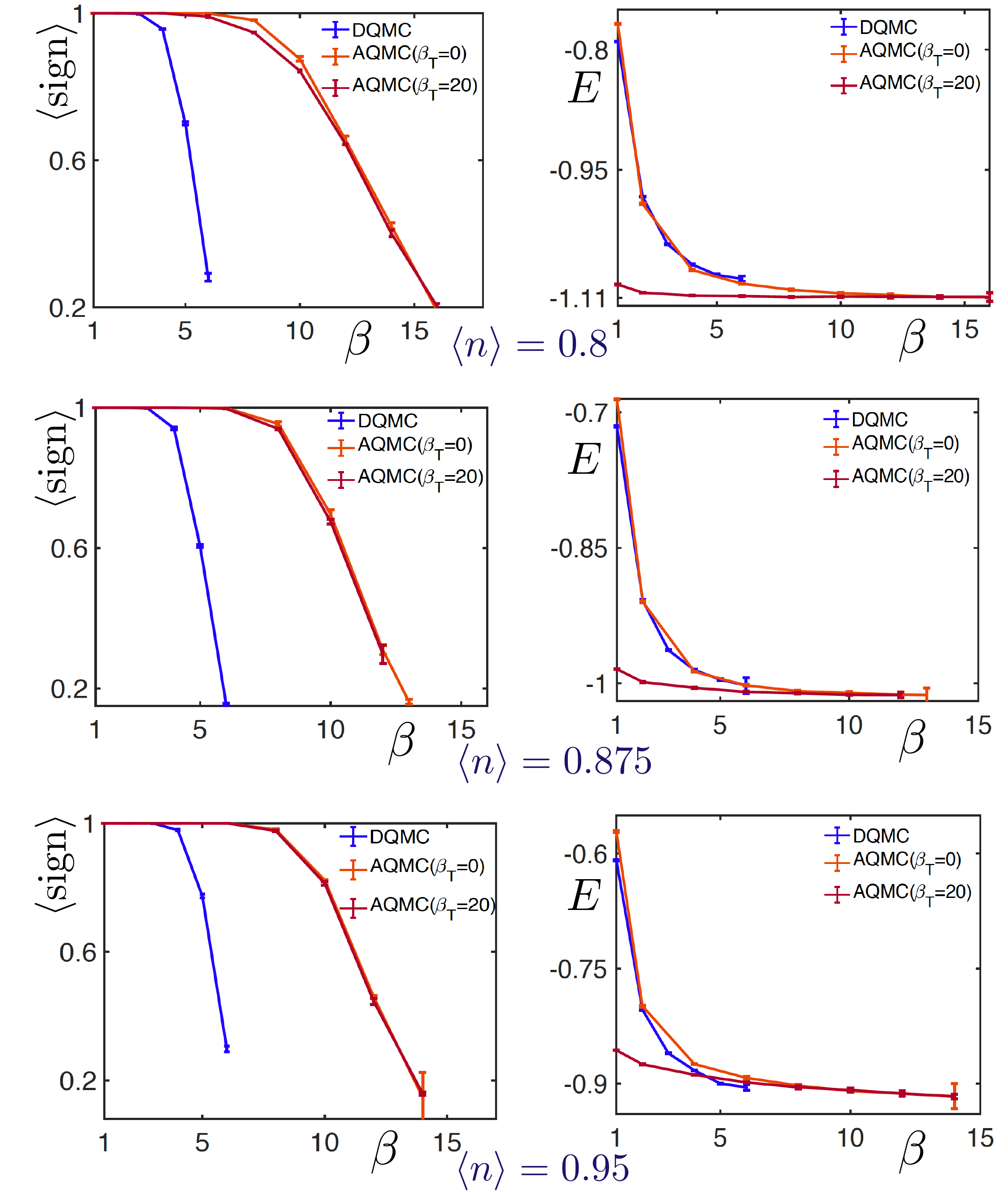}%
\caption{\raggedright Comparing the AQMC with DQMC for $<n>=0.8,0.875,0.9$ at $U=4$ for an $8\times8$ square geometry.}
\label{figS3}
\end{figure}

\begin{figure}[t]
\centering
\includegraphics[width=9 cm]{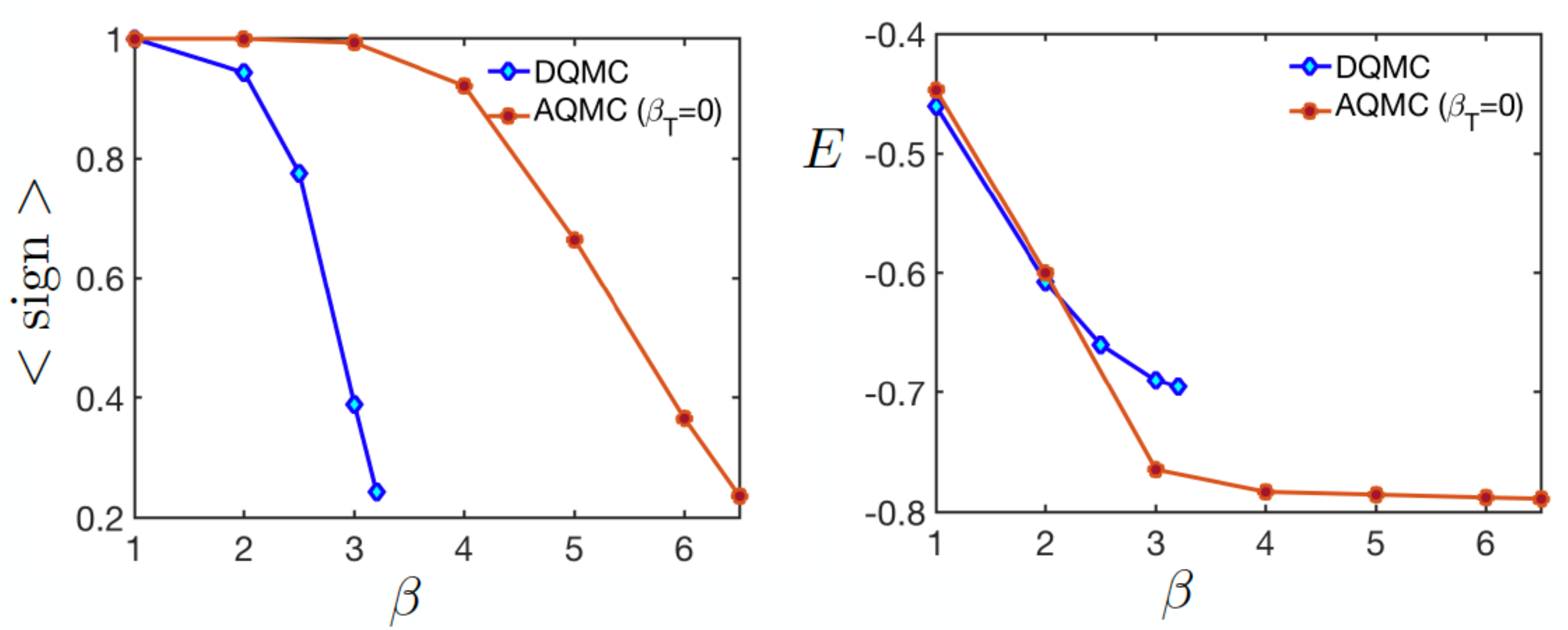}%
\caption{ Comparing the AQMC with DQMC for the Hubbard model above the Mott transition ($U=6$) at half filling on an $8\times8$ triangular geometry.}
\label{figtri}
\end{figure}

In Fig. \ref{figtri}, we show that the triangular model which has geometric frustration can be studied by our approach while in the standard DQMC, it is not possible below $T=1/3$ (beyond $\beta = 3$). 

\section{3. Benchmarking AQMC}

\begin{figure}[t]
\centering
\setlength{\belowcaptionskip}{-5pt}
 \includegraphics[width=8.5cm]{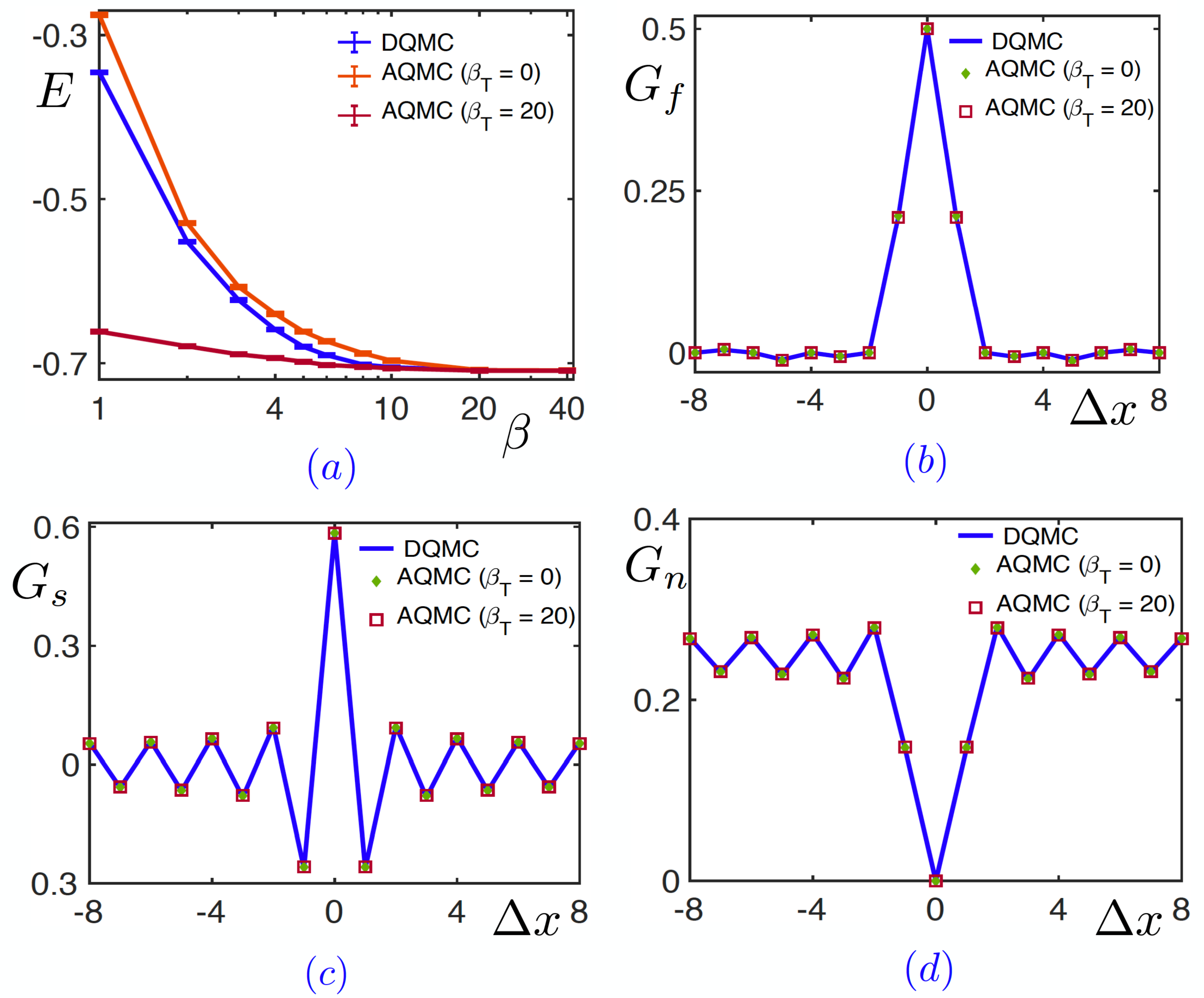}%
\caption{\raggedright Comparing the AQMC and DQMC for a $16\times 2$ system with $U=4$ at half filling. (a) Average energy obtained via DQMC, AQMC ($\beta_T=0$) and AQMC($\beta_T=20$) for several values of $\beta$. (b)-(d) Pair-wise comparison for the Green's function, spin-spin correlation function and pair distribution functions along the $x$ direction at $\beta = 40$.}
\label{fig1}
\end{figure}

$~$

In this section, we benchmark the AQMC algorithm by studying two cases with known exact results.
First, let us compare its performance and accuracy with the regular DQMC for the Hubbard model on the square lattice at half filling. Since this situation does not suffer from the sign problem, we can consider arbitrarily low temperatures and achieve the true ground-state properties within the regular DQMC. To this end, we consider a two-leg ladder with $\para{N_x,N_y}=(16,2)$ dimensions where electrons are coupled through $U=4$. We define the average energy per site as 
\bea
E = \frac{1}{N_s}\para{U\sum_{\bf i}\langle \hat{n}_{\bf i,\up}\hat{n}_{\bf i,\dn}\rangle - t_1\sum_{\langle \bf ij\rangle,\sigma} \langle \hat{c}_{\bf i,\sigma}^\dag \hat{c}_{\bf j,\sigma}  \rangle}, \label{Eq8}
\eea
where $N_s = N_xN_y$ denotes the number of sites. In Fig.~\ref{fig1}a, we have plotted the average energy at temperature $T=1/\beta$ for the DQMC for several temperatures down to $T = 1/40$ (in units of $t_1$). We then consider the same $\beta$ for the AQMC first for $\beta_T= 0$ and then for $\beta_T = 20$. As Fig.~\ref{fig1} clearly indicates, all three methods yield the same ground-state energy per site for large values of $\beta$ within the statistical error bars. For DQMC, we obtain $E = -0.7086 \pm 0.0003 (-0.7091 \pm 0.0003)$ at $\beta = 20$ ($\beta=40$). For $\beta_T = 0$, we achieve $E = -0.7085 \pm 0.0003$ for $\beta = 20$ and $-0.7088 \pm 0.0003$ for $\beta=40$. On the other hand for $\beta_T = 20$, we obtain $E = -0.7092 \pm 0.0003$ for both $\beta = 20$ and $\beta = 40$. Indeed, $\beta_T = 20$ is closely related to the projector QMC technique which is known to reach the ground-state for lower values of $\beta$ (compared to the regular DQMC). In Fig.~\ref{fig1}, besides the average energy, we have also compared the single-particle Green's function (for spin-up electrons), $G_{f}\para{\bf r} = \langle c_{\bf i,\up}^\dag c_{\bf i+r,\up} \rangle$, spin-spin correlation function, $G_{s}\para{\bf r} = \langle {\bf s}_{\bf i}. {\bf s}_{\bf i+r} \rangle$, and the pair distribution function for spin-up electrons, $G_{n}\para{\abs{\bf r}>0} = \langle {n}_{\bf i,\up} {n}_{\bf i+r,\up} \rangle$, and for ${\bf r} = \Delta x\hat{x}$. The results are fairly consistent.

$~$

\begin{figure}[t]
\centering
 \includegraphics[width=8.5cm]{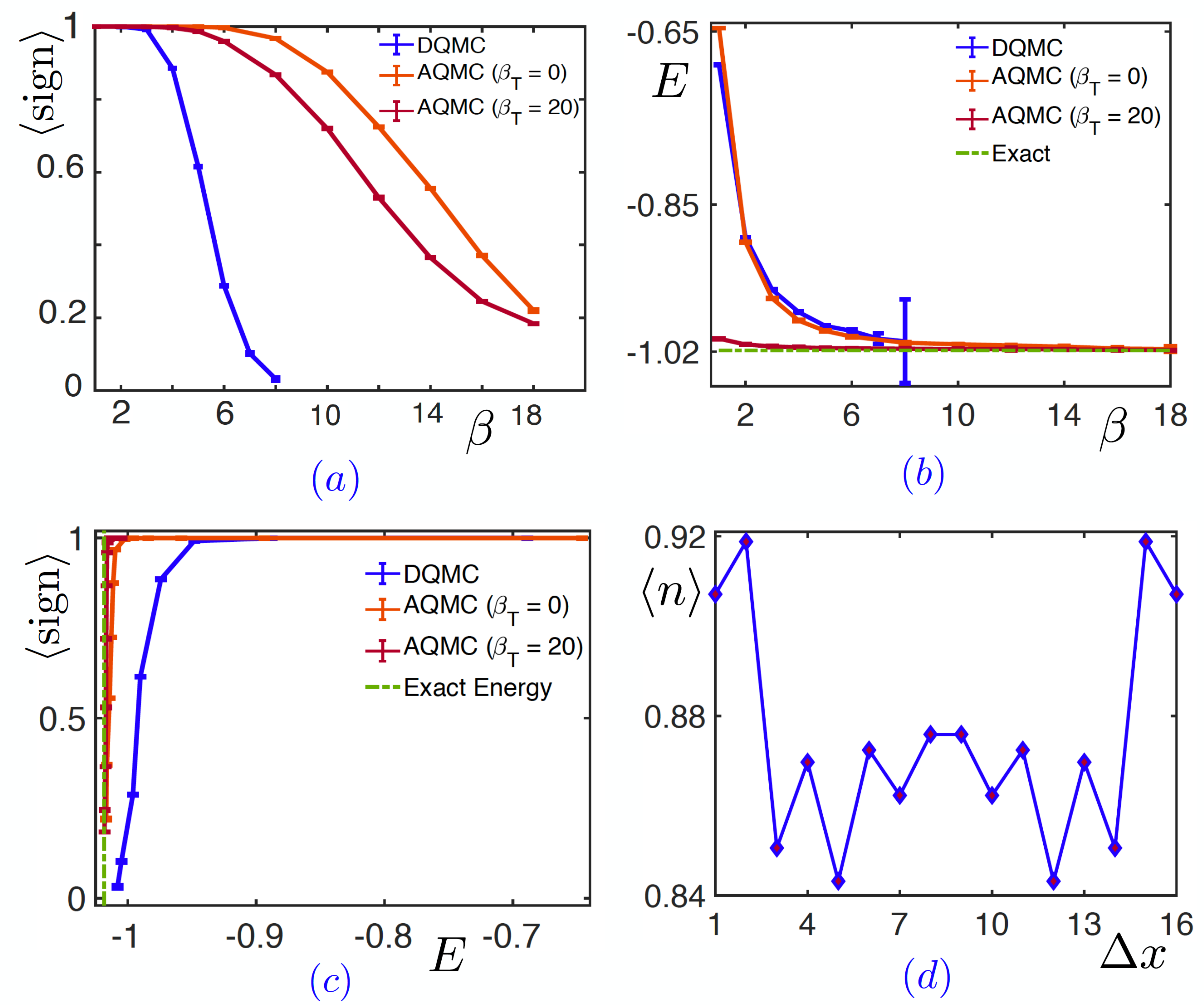}%
\caption{\raggedright (a)-(c) Comparing the results of AQMC with those obtained via DQMC and the exact result from DMRG's zero-truncation-error extrapolation for a $16\times 4$ cylindrical system at $p=1/8$ doping with $U=4$. (d) Local electron density achieved from AQMC with $\beta = 18$, and $\beta_T = 20$.}
\label{fig2}
\end{figure}

$~$

To further validate our method, we now focus on the doped Hubbard model which is known to suffer from the sign problem. Here, we consider $\para{N_x,N_y}=(16,4)$, $U=4$, and $p=1/8$ on a cylinder geometry. Fortunately, an accurate estimate of the ground-state energy for this problem is available from the DMRG approach. In Ref.~\cite{ehlers2017hybrid}, this and a number of other related problems have been studied by considering bond dimensions as large as $M = 35000$, and their finite $M$ results are then extrapolated to $M = \infty$ (more precisely the zero truncation error limit) to extract the true ground-state energy. Although for this particular case, the authors have only reported their infinite $M$ extrapolation results with $E_g = -1.01841$ for the ground-state energy per site, from their other available extrapolations, we believe that for $16\times 4$ system the average truncation error is about $10^{-5}$ for $M=35$k. Therefore DMRG's actual estimate at this bond dimension is indeed approximately $E= -1.0175 \pm 0.0005$. Note that such an enormous bond dimension will require weeks of CPU-time on state-of-the-art supercomputers. In contrast, we achieved $E= -1.0182 \pm 0.0003$ using the AQMC (for $\beta = 18$, $\beta_T = 20$) within a few hours on a personal computer.

In Fig.~\ref{fig2}a, we have plotted the average sign at a given $\beta$ for the regular DQMC, as well as the AQMC for both $\beta_T = 0$, and $\beta_T = 20$. As promised earlier, the average sign is significantly higher in the AQMC. In Fig.~\ref{fig2}b, the average energy as a function of $\beta$ is presented for all three simulations. Again, we can verify that the AQMC, especially for $\beta_T = 20$, almost reaches the true ground-state energy (corroborated via DMRG results extrapolated to $M=\infty$~\cite{ehlers2017hybrid}) within the statistical error bar of our samplings. On the other hand, due to the severity of the sign problem, for the regular DQMC, we could not draw any meaningful conclusion beyond $\beta = 7$ where $\langle {\rm sign}\rangle \approx 0.1$ and $E \approx -1.007 \pm 0.005$.
 Another useful plot is the average sign as a function of average energy. In the AQMC, we can probe ground-state properties before the sign problem becomes uncontrollable. Finally, we have plotted the electron density as a function of $x$ which is similar to Fig. 6a of Ref.~\cite{ehlers2017hybrid} (though they have plotted it for a $32\times 4$ cylinder).

\begin{figure}[t]
\centering
\includegraphics[width=8 cm]{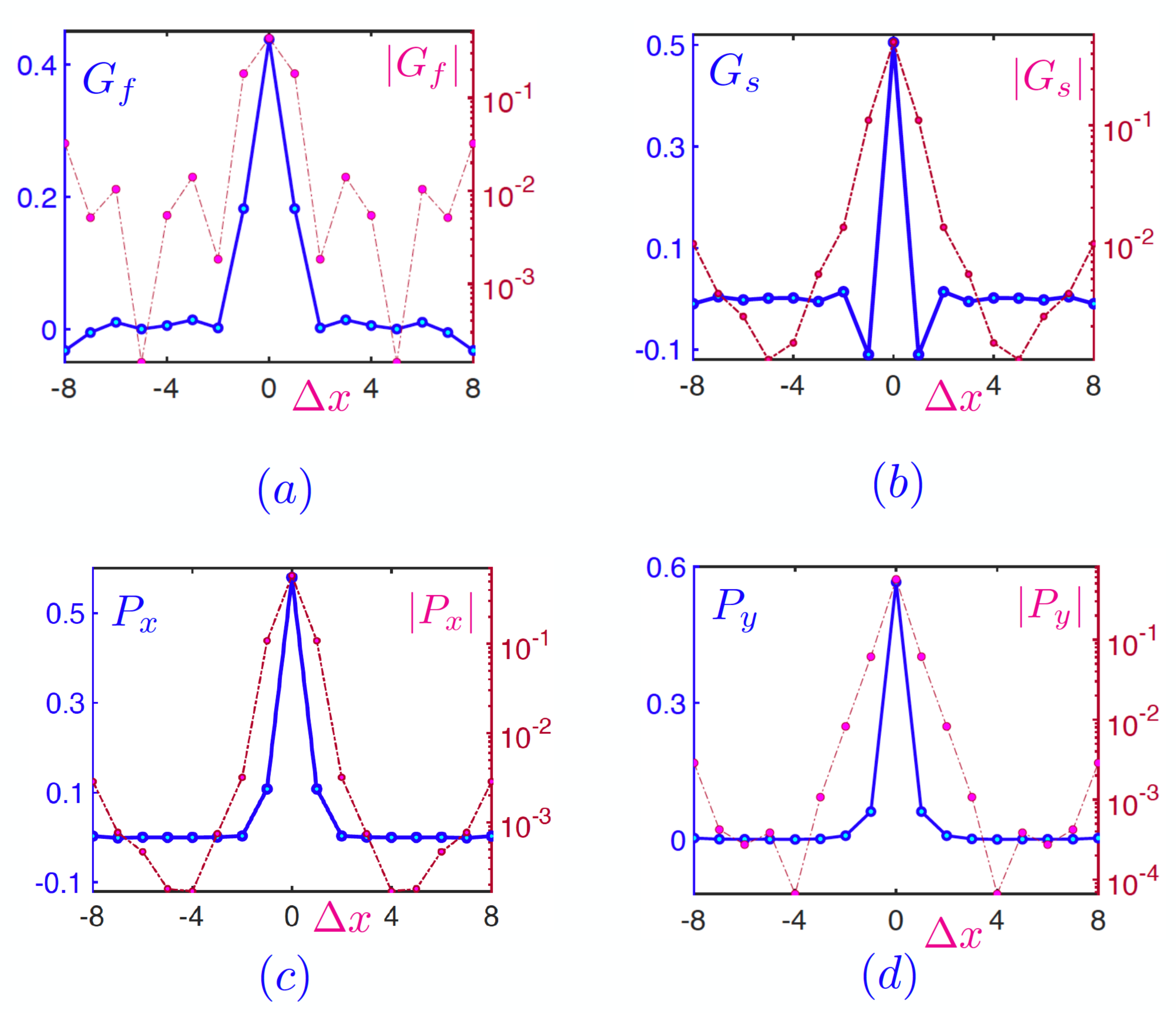}%
\caption{\raggedright Various correlation functions along the $x$ axis obtained via AQMC with $\beta=12$, and $\beta_T = 20$ for the $16\times 8$ sample at $p=1/8$ and $U=4$: (a) Green's function of spin up electrons, (b) Spin-spin correlations function, (c) and (d) Superconducting (pair-pair) correlation function for $x$ and $y$ bonds, respectively.}
\label{figS4}
\end{figure}

\begin{figure}[t]
\centering
\includegraphics[width=8 cm]{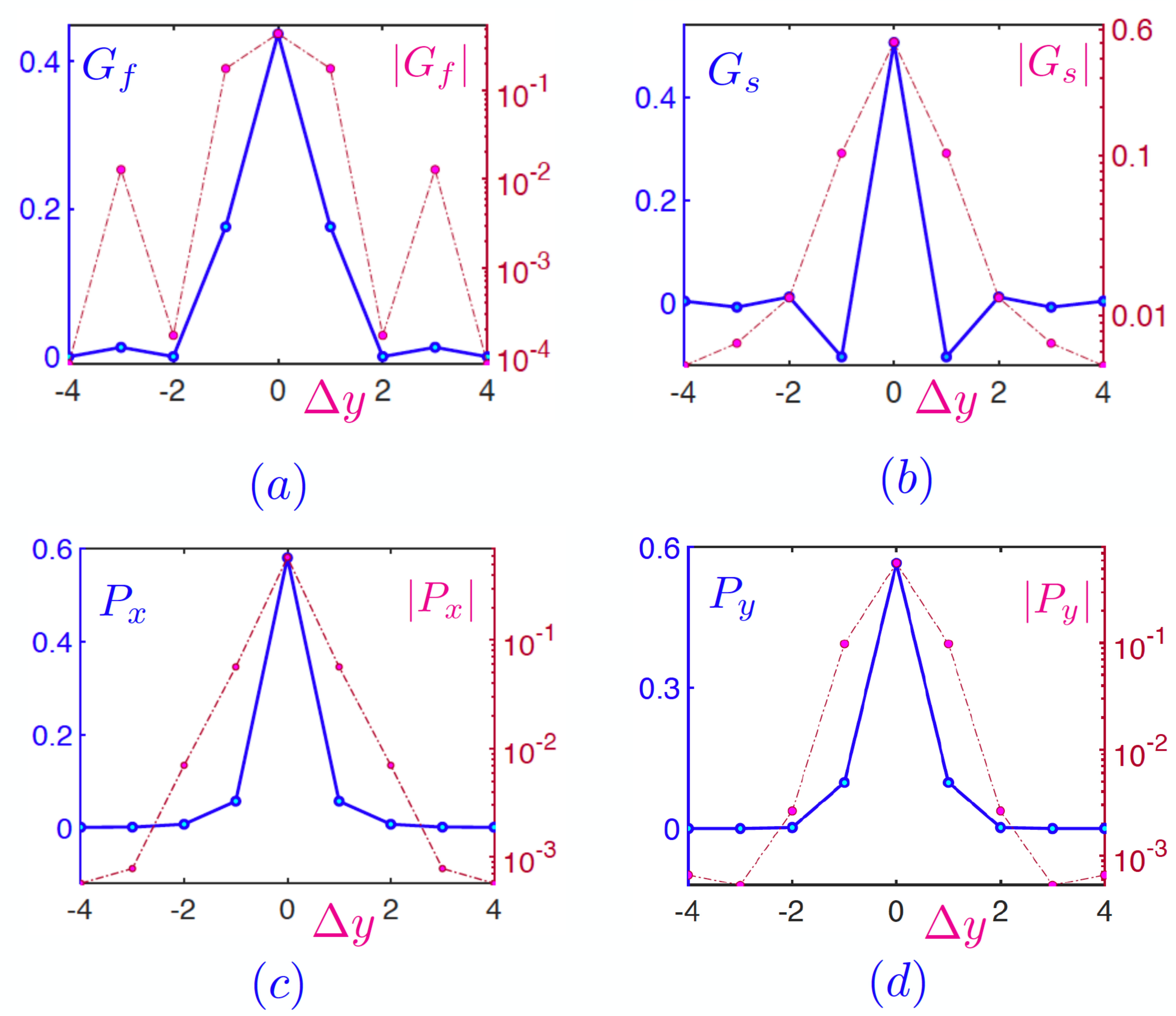}%
\caption{\raggedright Various correlation functions along the $y$ axis obtained via AQMC with $\beta=12$, and $\beta_T = 20$ for the $16\times 8$ sample at $p=1/8$ and $U=4$: (a) Green's function of spin up electrons, (b) Spin-spin correlations function, (c) and (d) Superconducting (pair-pair) correlation function for $x$ and $y$ bonds, respectively.}
\label{figS5}
\end{figure}

$~$

\section{4. More correlation functions for $16\times 8$ system}
We here provide more details about the $U=4$, $16\times 8$ system at $p=1/8$ doping level. For the two-point correlations, we have considered  $\beta = 12$, $\beta_T = 20$ with about seven million spacetime sweeps in total.  We have measured the spin-resolved Green's function ($G_{f}\para{\bf r}$), spin-spin ($G_{s}\para{\bf r}$) and also the superconducting (pair-pair) correlation functions ($P_{x}\para{\bf r}$ and $P_{y}\para{\bf r}$). The pair-pair correlations are related to the following two pairing fields: $\hat{\Delta}_{x}\para{\bf i} \equiv \hat{c}_{\bf i , \up} \hat{c}_{{\bf i} + \hat{x}, \dn}- \hat{c}_{\bf i , \dn} \hat{c}_{{\bf i} + \hat{x}, \up}$ and $\hat{\Delta}_{y}\para{\bf i} \equiv \hat{c}_{\bf i , \up} \hat{c}_{{\bf i} + \hat{y}, \dn}-\hat{c}_{\bf i , \dn} \hat{c}_{{\bf i} + \hat{y}, \up}$. Accordingly, we define $P_{x}\para{\bf r} = \langle \Delta_{x}\para{\bf i}^\dag \Delta_{x}\para{\bf i + r}\rangle$, $P_{y}\para{\bf r} = \langle \Delta_{y}\para{\bf i}^\dag \Delta_{y}\para{\bf i + r}\rangle$. We employ the translation symmetry and  average over all possible $\bf i$'s.  Figs.~\ref{figS4} and \ref{figS5} summarize our results. 

Our spin-spin correlation function clearly points towards the stripe order formation. As Fig.~\ref{figS4}b shows, we evidence a $\pi$ phase shift in the correlations after $\abs{\Delta x}=4$. 
While for $\abs{\Delta x}\leq 4$, spins with odd (even) distances along the $x$ axis (and with identical $y$'s) have negative (positive) correlations, we observe the opposite behavior for $4 < \abs{\Delta x}\leq 8$. 
Another exotic feature of the correlations is that they are all enhanced near $\abs{\Delta x} = 8$, and grow with distance (rather than decay) for $5<\abs{\Delta x}\leq 8$. 
In contrast, correlation functions along the $y$ axis (c.f., Fig.~\ref{figS5}) behave normally and as expected.

$~$

\section{5. Emergence of $\nu =1/2$ Laughline state in the correlated Chern insulators}

Now, we would like to elucidate how the topological order can emerge from interacting Chern insulators. Our approach is closely related to the parton construction of fractional quantum Hall states. First, let us recall that the low energy description of a non-interacting Chern insulator is given by a Chern-Simons gauge theory after integrating out the fermionic degrees of freedom. Thus, we have:
\bea
&&\mathcal{L}_{\rm tot} = \mathcal{L}_{\up} + \mathcal{L}_{\dn},\cr
&&\mathcal{L}_{\sigma = \up,\dn} = \frac{1}{4\pi} \epsilon^{\mu\nu \lambda}a_{\mu,\sigma} \partial_{\nu} a_{\lambda,\sigma} - \frac{e}{2\pi} \epsilon^{\mu\nu \lambda} a_{\mu,\sigma} \partial_{\nu} A_{\lambda,\sigma},
\eea
where $\bf A$ is the electron-magnetic (external/probe) gauge field, and $\bf a_{\sigma}$ is related to the current of spin-$\sigma$ electrons as follows:
\be
j^{\mu}_{\sigma} = \frac{1}{2\pi} \epsilon^{\mu\nu \lambda} \partial_{\nu} a_{\lambda,\sigma}.
\ee
The above Chern-Simons theory describes a system with $U(1)\times U(1)$ symmetry, where the first $U(1)$ denotes the charge sector, and the second one the spin sector. The above model does not have topological order, meaning that it has a unique ground-state on the torus geometry (periodic system), and contains only trivial (fermionic or bosonic) excitations and does not support any anyon excitation.
Now, let us consider a strong onsite attraction between electrons with opposite spins. Such a strong attraction will force the density and current of the two species to become equal at low temperatures. This constraint can be enforced via imposing $ a_{\mu,\up}\para{\bf r} = a_{\mu,\dn}\para{\bf r} = b_{\mu}$. In other words, the interaction gaps out the spin sector via Higgsing the spin gauge field: $a_{s,\mu}\para{\bf r} \equiv a_{\mu,\up}\para{\bf r}-a_{\mu,\dn}\para{\bf r}$ that can for example be achieved through adding $m_s^2 a_{s,\mu}a_{s}^{\mu}$ term to the effective Lagrangian. Implementing the $a_{\mu,\up}\para{\bf r} = a_{\mu,\dn}\para{\bf r} = b_{\mu}$ constraint in the above effective Lagrangian, we arrive at:

\bea
&&\mathcal{L}_{\rm eff} = \frac{2}{4\pi} \epsilon^{\mu\nu \lambda}  b_{\mu} \partial_{\nu} b_{\lambda} - \frac{2e}{2\pi} \epsilon^{\mu\nu \lambda} b_{\mu,\sigma} \partial_{\nu} A_{\lambda,\sigma},
\eea

The above effective $U(1)_2$ Chern-Simons theory describes a liquid of charge $q=2e$ constituents (doublons) with a on-trivial topological order. Their Hall conductivity is $\sigma_{xy} = \frac{1}{2}\frac{q^2}{h}$, indicating the effective filling fraction (of doublons), $\nu$, equals $1/2$. Moreover, the ground-state of the above theory has a two-fold degeneracy on the torus geometry, and contains only two distinct excitations, namely: trivial (bosnic) and semionic excitations. Hence, all other excitations (including electrons) are confined (i.e., they are highly massive and absent from the effective low energy description).

In our QMC study of this problem, we adopt the staggered $\pi$-flux model which is a Chern insulator and has the following tight binding form:
\bea
H_0 = && \sum_{\sigma = \up,\dn}\sum_{\bf k} a_{\bf k}\para{c_{A,\bf k,\sigma}^\dag c_{A,\bf k,\sigma}-c_{A,\bf k,\sigma}^\dag c_{A,\bf k,\sigma}}\cr
+ && \sum_{\sigma = \up,\dn}\sum_{\bf k} \para{b_{\bf k}c_{A,\bf k,\sigma}^\dag c_{B,\bf k,\sigma}+h.c.},
\eea
where
\be
a_{\bf k} = 2t_2 \para{\cos~{k_x}-\cos~{k_y}},
\ee
and
\be
b_{\bf k} = t_1 e^{-i\pi/4} \left[1+e^{i\para{k_y-k_x}}\right] + t_1e^{i\pi/4} \left[e^{-ik_x}+e^{ik_y}\right]
\ee
To achieve the maximum flatness of the band-structure, we consider $t_2/t_1=\sqrt{2}$ (and $t_1=1$).

\begin{figure}[t]
\centering
\includegraphics[width=9 cm]{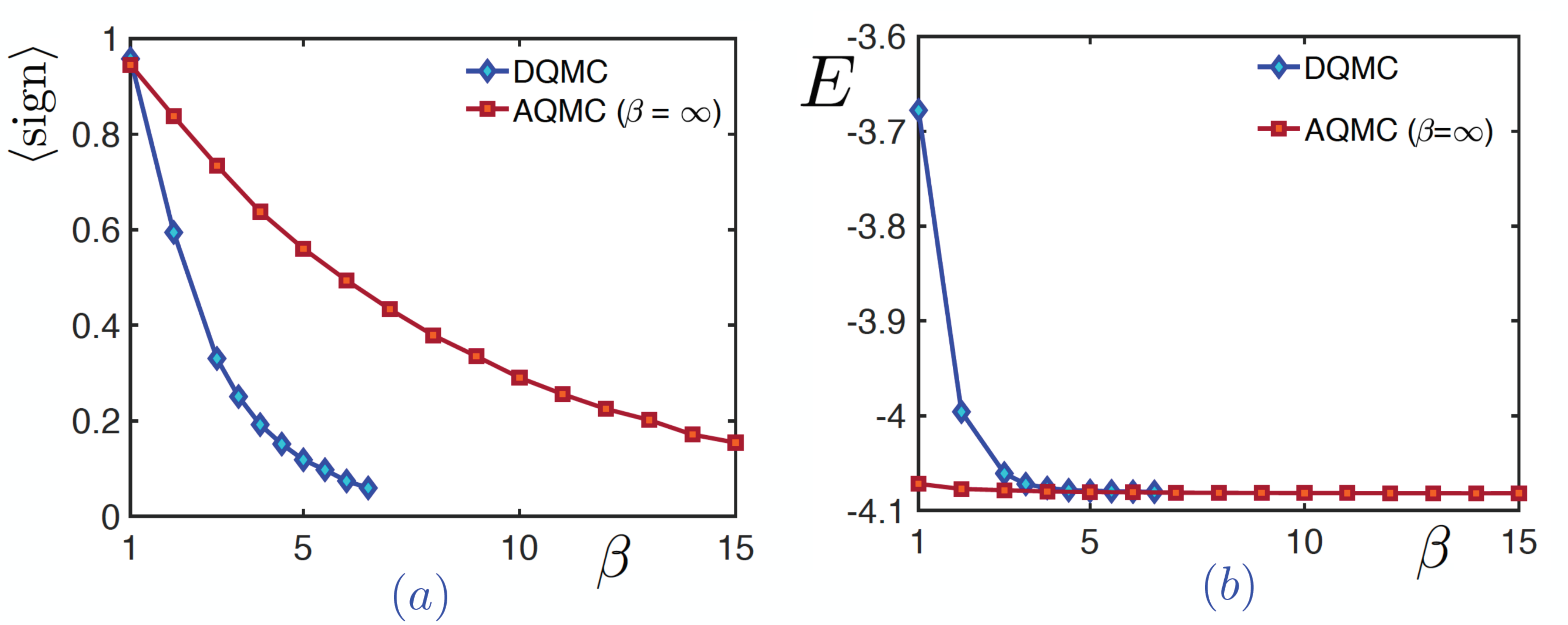}%
\caption{\raggedright Comparing the AQMC with DQMC for the correlated Chern insulator model for $U=-6$ on an $8\times8$ cylinder were we impose a periodic (open) boundary condition along the $x$ ($y$) direction. The AQMC yields a significantly boosted average sign and a slightly lower energy estimate, which plateaus and must correspond to its true zero temperature value.}
\label{figS10}
\end{figure}

We fill the lower band of the model with spin up and spin down electrons, respectively to achieve the half filling. Next, we couple electrons via an onsite Hubbard attraction with $U=-6$. Although our model is at half filling and subject to an attractive interaction, it suffers from the sign problem due to the lack of time-reversal symmetry (which is required to guarantee the absence of sign problem in the regular undoped Hubbard model). In Fig. \ref{figS10}, we present the average sign and energy of this model on an $8\times8$ cylinder.

\begin{widetext}

\section{6. MATLAB implementation of AQMC}
Below, we provide a simple MATLAB implementation of the AQMC as well as DQMC algorithms. For $\beta_T=0$, this is an efficient and optimized code whose performance and speed are comparable to C++ and Fortran implementations. For $\beta_T > 0$, to avoid unnecessary programming complications, we break $\beta_T$ into $N_T = \beta_T / \Delta \tau$ imaginary time steps, instead of a single imaginary time step (i.e., $N_T = 1$, $\Delta \tau = \beta_T$). The other choice, namely $\Delta \tau_T = \beta_T$ will requires several modifications to the code, but instead will reduce the CPU time significantly.

To run this code, copy the main script (marked in blue) and every function (marked in red) in separate MATLAB files. Next, run the main script. In the main script, the user can tune several physical parameters. For example, the onsite Hubbard interaction strength $U$, the inverse temperature $\beta = T^{-1}$, chemical potential $\mu$, the dimensions of the system along the $x$ ($y$) direction, $N_x$ ($N_y$), and the boundary condition along the $x$ ($y$) axis can be altered easily. A number of other parameters can be tuned as well. For example, the (imaginary) time steps $\Delta\tau$, $\beta_T$, $\mathscr{U}_{\rm max}$, etc can be determined by the user. For the regular DQMC, we just need to set {\it swtch$\_$adiabatic=0}, and $\beta_T = 0$. For systems with more than about $50$ sites, our code automatically employs the delayed update method (which is based on the Sherman-Morrison-Woodbury identity) instead of the standard Sherman-Morrison update rule for the Metropolis algorithm. It is significantly faster for large systems.

This code can be easily parallelized by running the Markov chains in parallel as they are entirely independent. This can be achieved through the {\it parfor} command in the {\it func$\_$QMC} function, and the user needs to uncomment the {\it parpool()} command (or alternatively uses {\it parpool(nw)}, where {\it nw} denotes the number of workers (CPUs) available on the computer) at the beginning of the main script.

For the sake of convenience, our code computes the following two-point correlation functions as well: $\langle c_{\bf i,\sigma}^\dag c_{\bf j,\sigma}\rangle$, $\langle s_{z,{\bf i}} s_{z,{\bf j}}\rangle$, $\langle s_{x,{\bf i}} s_{x,{\bf j}}\rangle$, and $\langle n_{{\bf i},\up} n_{{\bf j},\up}\rangle$, provided {\it swtch$\_$correlation = 1}.

{\color{blue}
\begin{tiny}
\begin{verbatim}
%%%%%%%%%%%%%%%%%%%%%%%%%%%%%%%%%%%%%%%%%%%%%%%%%%%%%%%%%%%%%%%%%%%%%%%%%%%%%%%%
%===============================================================================
%              Adiabatic & Determinent Quantum Monte Carlo
%        Simple Matlab code for QMC study of 1D & 2D Hubbard model
%                            Main script
%===============================================================================

clc;
clear;
close all;
warning 'off';

% parfor(); % only if parallelization is desired

%===============================================================================
%               QMC parameters
%===============================================================================

U = 4.0; % Hubbard interaction strength
mu = 0.0; % chemical potential (@ half-filling: mu =0)
n_x = 10; % # of sites along x direction
n_y = 2;  % # of sites along y direction
bnd_x = 1; % boundary condition along x:  0 --> open  1--> periodic
bnd_y = 1; % boundary condition along y:  0 --> open  1--> periodic
beta = 4; % inverse temperature:    beta = 1/T
dtau = 1/10; % imaginary time steps
dk = 10; % stratification period (default = 10)

n_measurement = 5000; % number of measurement sweeps
n_warmup = round(n_measurement/5); % number of warmup (burn-in) sweeps
n_markov = 6; % number of independent markov chains
% 0 --> only energy is computed
% 1 --> 2-point correlations are also computed
swtch_correlation = 1; 

% Adiabatic QMC parameters
swtch_adiabatic = 0; % 0--> regular DQMC 1--> adiabatic QMC
beta_trial = 4; % inverse temperature of trial density matrix
u_max = 4; % u(t) will eventually reach u_max (U) at tau = beta/2
if swtch_adiabatic == 0
    u_max = U;
end
%===============================================================================
%               Hopping amplitudes
%===============================================================================

t_x = 1; % NN hopping along x axis
t_y = t_x; % NN hopping along y axis
t_xy = 0; % NNN hopping
hopping = hopping_matrix(t_x,t_y,t_xy,n_x,n_y,bnd_x,bnd_y);
% adding chemical potential term to enforce the desired electron density
hopping = hopping - mu*diag(ones(1,n_x*n_y)); 


%===============================================================================
%               QMC algorithm
%===============================================================================

tic;
% func_QMC function contains the QMC algorithm's main steps
[par_func_tmp,correlations_tmp] = func_QMC(swtch_adiabatic,beta_trial,u_max,...
      hopping,beta,n_x,n_y,dtau,dk,n_markov,n_warmup,n_measurement,swtch_correlation);
elapsed_time = toc

%===============================================================================
%               Measurements
%===============================================================================

% working on outputs of func_QMC to generate the desired expectation values 
% in expectation_values_computation function
[correlations,measurements] = exp_val(par_func_tmp,correlations_tmp,...
      n_markov,U,mu,n_x,n_y,t_x,t_y,t_xy,bnd_x,bnd_y);


%===============================================================================
%               Plots of two-point correlation functions
%===============================================================================

if swtch_correlation == 1
    x_axis = -round(n_x/2):n_x-round(n_x/2); % distance
    % spin-spin correlation (z component)
    figure(1);
    cor = circshift(correlations{4}{1},-round(n_x/2),2);
    plot(x_axis, [cor,cor(1)],'-o','LineWidth',2);
    xlabel('distance');
    ylabel('G_{s,zz}');
    title('spin-spin correlation (z component)');

    % spin-spin correlation (x component)
    figure(2);
    corr = circshift(correlations{4}{2},-round(n_x/2),2);
    plot(x_axis, [corr,corr(1)],'-d','LineWidth',2);
    xlabel('distance');
    ylabel('G_{s,xx}');
    title('spin-spin correlation (x component)');

    % density-density correlation (for spin up electrons)
    figure(3);
    cor3 = circshift(correlations{4}{3}(1,:),-round(n_x/2),2);
    plot(x_axis,[cor3,cor3(1)] ,'-o','LineWidth',2);
    xlabel('distance');
    ylabel('G_{n}');
    title('pair distribution function');

end
%%%%%%%%%%%%%%%%%%%%%%%%%%%%%%%%%%%%%%%%%%%%%%%%%%%%%%%%%%%%%%%%%%%%%%%%%%%%%%%%
\end{verbatim}
\end{tiny}
}

{\color{purple}
\begin{tiny}
\begin{verbatim}
%%%%%%%%%%%%%%%%%%%%%%%%%%%%%%%%%%%%%%%%%%%%%%%%%%%%%%%%%%%%%%%%%%%%%%%%%%%%%%%%
function hopping = hopping_matrix(t_x,t_y,t_xy,n_x,n_y,bnd_x,bnd_y)

% constructing the matrix of hopping amplitudes for a 2D square lattice
dx = 1;
rx_max = dx; % range of hopping along x
if n_x == 1
    rx_max = 0;
end
dy = 1;
ry_max = dy; % range of hopping along y
if n_y == 1
    ry_max = 0;
end


% hopping amplitudes from sit i to its nearest and next nearest neighbors
T= -[t_xy     t_x   t_xy;
     t_y      0     t_y;
     t_xy     t_x   t_xy];

% hopping matrix
hopping = zeros(n_x*n_y);
for i = 1 : n_x
    for j = 1 : n_y
        % index of site r = (i,j)
        r = (i-1)*n_y + j; 
        for rx = -rx_max:rx_max
            ii = i + rx;
            if bnd_x == 1
                ii = 1+mod(ii-1,n_x);
            end
            for ry = -ry_max:ry_max
                jj = j + ry;
                if bnd_y == 1
                    jj = 1+mod(jj-1,n_y);
                end

                % index of site rr = (ii,jj)
                rr = (ii-1)*n_y + jj; 

                if ii>0 && ii <=n_x
                    if jj >0 && jj <=n_y
                        hopping(r,rr) = T(rx+dx+1,ry+dy+1);
                        hopping(rr,r) = hopping(r,rr)';
                    end
                end
            end

        end
    end
end
hopping = (hopping + hopping')/2;

end
%%%%%%%%%%%%%%%%%%%%%%%%%%%%%%%%%%%%%%%%%%%%%%%%%%%%%%%%%%%%%%%%%%%%%%%%%%%%%%%%
\end{verbatim}
\end{tiny}

\begin{tiny}
\begin{verbatim}
%%%%%%%%%%%%%%%%%%%%%%%%%%%%%%%%%%%%%%%%%%%%%%%%%%%%%%%%%%%%%%%%%%%%%%%%%%%%%%%%
function [par_func_tmp,correlations_tmp]=func_QMC(swtch_adiabatic,beta_trial,...
      u_max,hopping,beta,n_x,n_y,dtau,dk,n_markov,n_warmup,n_measurement,swtch_corr)

warning 'off';

n_s = n_x*n_y; % total # of sites

n_tau = max(dk,2*round((beta+beta_trial)/(2*dtau))); % # of time steps
dtau = (beta + beta_trial)/n_tau; % time steps
L = round(beta/(2*dtau)); % # of time steps for the evolution period

u_vec = zeros(n_tau,1);
for l = 1:n_tau
    if l*dtau <= beta/2
        u_vec(l) = u_max*(1 - swtch_adiabatic*(l-1)/L);
    elseif l*dtau > (beta/2+beta_trial)
        u_vec(l) = u_max*(1 - swtch_adiabatic*(n_tau-l)/L);
    end
end
lambda = acosh(exp(u_vec*dtau/2));

%==========================================================================
% computing expm(-KdT), expm(KdT), expm(-KdT/2), & expm(KdT/2) where K is
% the kinetic term (hopping matrix)
%==========================================================================

[U0,D0] = eig(hopping);
D0_d = diag(D0);
B = U0*diag(exp(-dtau*D0_d))*U0';
B_root = U0*diag(exp(-dtau*D0_d/2))*U0';
inv_B = U0*diag(exp(dtau*D0_d))*U0';
inv_B_root = U0*diag(exp(dtau*D0_d/2))*U0';

%==========================================================================

num_samples_avg = zeros(n_markov,1);
sign_avg = zeros(n_markov,1);
GF_up_avg = cell(n_markov,1);
GF_dn_avg = cell(n_markov,1);
n_up_dn_avg = cell(n_markov,1);
two_point_corr_avg = cell(n_markov,1);

%==========================================================================

parfor count = 1:n_markov    
    
    %========================================================================
    
    % to be used for Green's function evaluation and space-time wrap
    k_vec = zeros(1,n_tau); 
    k_vec(dk:dk:end) = 1;
    k_vec(end) = 1;

    % Initalization:
    
    num_samples = 0;
    sign_Z = 0;
    
    two_point_corr = cell(1,3);
    n_up_dn = 0;
    GF_up = 0;
    GF_dn = 0;
    for m = 1 : 3
        two_point_corr{m} = 0 ;
    end
    % to be used for 2-point correlations
    range_corr_x = 1:n_x;
    range_corr_y = 1:n_y;

    %========================================================================
    
    % random initialization Hubbard-Stratonovich field config.
    s = (2*randi([0,1],n_tau,n_s)-1); 
    % h_up = lambda*s, h_dn = -h_up = -lambda*s
    h = diag(lambda)*s; 
    
    % computing corresponding Green's functions for up and down spin electrons
    % computing B_s(l) := exp(-KdT)exp(-h_s(l)) for all l = 1:n_T  
    B_up_cell = cluster(B,inv_B,h,n_tau);    
    B_dn_cell = cluster(B,inv_B,-h,n_tau); 
    % Using QR decomposition to stabilize B_L ... B_1 
    % to compute G = inv(Id + B_L ... B_1)  for B = B_up or B_dn 
    [G_up,G_dn,log_det_G,sign_det_G] = initial_eval(B_up_cell,B_dn_cell,k_vec,n_tau);
    
    % the sign of various HS field configs is needed for computing 
    % expectation values
    sign_Z_accumulated = sign_det_G;
    
    %========================================================================
    
    % space-time sweeps:
    for i_swp = 1:(n_warmup+n_measurement)
        for l = 1: n_tau
            % Green's function update upon Hubbard-Stratonovich fields flip 
            % for large systems delayed update (Sherman-Morrison-Woodbury) is faster
            % than Sherman-Morrison update method
            if n_s > 50 
                delayed_period = 32; 
                [G_up,G_dn,h,sign_Z_accumulated_tmp] = delayed_update(G_up,G_dn,h,n_s,delayed_period);
            else
                [G_up,G_dn,h,sign_Z_accumulated_tmp] = Sher_Mor(G_up,G_dn,h,n_s);
            end
            % sign_current = sign_prev*sign_tmp
            sign_Z_accumulated = sign_Z_accumulated*sign_Z_accumulated_tmp;
            % space time wrap up by one time-step
            h = circshift(h,-1,1);
            k_vec = circshift(k_vec,-1,2);
            
            if k_vec(l) == 1
                % reevaluating the Green's function from scratch
                B_up_cell = cluster(B,inv_B,h,n_tau);
                B_dn_cell = cluster(B,inv_B,-h,n_tau);
                [G_up,G_dn,~,~] = initial_eval(B_up_cell,B_dn_cell,k_vec,n_tau);
                
            else
                % spacetime wrapping using already existing Green's functions
                h_tmp = h(n_tau,:);
                
                % spin up
                B_up_prev = B*diag(exp(h_tmp));
                inv_B_up_prev = diag(exp(-h_tmp))*inv_B;
                G_up = B_up_prev*G_up*inv_B_up_prev;
                
                % spin dn
                B_dn_prev = B*diag(exp(-h_tmp));
                inv_B_dn_prev = diag(exp(h_tmp))*inv_B;
                G_dn = B_dn_prev*G_dn*inv_B_dn_prev;
            end
            
        end
        
        %==============================================
        % Measurements for the current HS field config.
        %==============================================
        
        if  i_swp > n_warmup
            num_samples = num_samples + 1;

            sign_Z = sign_Z + sign_Z_accumulated;

            % inv_B_root*G_up*B_root & inv_B_root*G_dn*B_root
            % instead of G_up and G_dn are used to employ the
            % 2nd order Trotter-Suzuki decomposition
            GF_up_tmp = eye(n_s) - inv_B_root*G_up*B_root;
            GF_dn_tmp = eye(n_s) - inv_B_root*G_dn*B_root;

            % density profile for the current HS fields
            n_up_tmp = diag(GF_up_tmp);
            n_dn_tmp = diag(GF_dn_tmp);

            % spin and total density for the current HS fields
            sz_tmp  = (n_up_tmp-n_dn_tmp)/2;
            rho_tmp = (n_up_tmp+n_dn_tmp)/2;

            % to compute average GF and <n_up n_dn>
            GF_up = GF_up + GF_up_tmp*sign_Z_accumulated;
            GF_dn = GF_dn + GF_dn_tmp*sign_Z_accumulated;
            n_up_dn = n_up_dn + n_up_tmp.*n_dn_tmp*sign_Z_accumulated;

            % two-point correlation functions: spin-spin
            % and density-density correlation functions along x axis
            % namely: <O(x0,y0)O(x0+x,y0)>
            if swtch_corr == 1
                corr_tmp = cell(1,3);

                corr_0 = cell(1,3);
                for m = 1 : 3
                    corr_0{m} = 0;
                end
                count_tmp = 0;
                for ix_1 = range_corr_x
                    for iy_1 = range_corr_y
                        r_1 = (ix_1-1)*n_y + iy_1; % first coordinate of the correlation function
                        r_2 = n_y*[(ix_1:n_x) (1:(ix_1-1))] - n_y + iy_1; % 2nd coordinate of the correlation function

                        %  <sz(r_1)sz(rr_2)> computation
                        corr_tmp{1} = (sz_tmp(r_1)*sz_tmp(r_2)).'- (GF_up_tmp(r_1,r_2).*GF_up_tmp(r_2,r_1).')/4 ...
                               -(GF_dn_tmp(r_1,r_2).*(GF_dn_tmp(r_2,r_1)).')/4;

                        corr_tmp{1}(1) = -n_up_tmp(r_1)*n_dn_tmp(r_1)/2 + (rho_tmp(r_1))/2;

                        %  <sx(r_1)sx(r_2)> computation
                        corr_tmp{2} = -GF_up_tmp(r_1,r_2).*(GF_dn_tmp(r_2,r_1).')/4 - GF_dn_tmp(r_1,r_2).*(GF_up_tmp(r_2,r_1).')/4;

                        corr_tmp{2}(1) = -n_up_tmp(r_1)*n_dn_tmp(r_1)/2 +(rho_tmp(r_1))/2;

                        %  <n_{up}(r_1)n_{up}(r_2)> computation
                        corr_tmp{3} =  n_up_tmp(r_1)*n_up_tmp(r_2).'- (GF_up_tmp(r_1,r_2).*GF_up_tmp(r_2,r_1).');
                        corr_tmp{3}(1) = 0; 


                        for m = 1 : 3
                            corr_0{m} = corr_0{m} + corr_tmp{m};
                        end
                        count_tmp = count_tmp + 1;

                    end
                end
                for m = 1 : 3
                    corr_0{m} = corr_0{m}/count_tmp;
                end

                for m = 1 : 3
                    two_point_corr{m} = two_point_corr{m} + corr_0{m}*sign_Z_accumulated;
                end

            end
        end
    end
    
    % <O> =  <Osign(z)>/<sign(z)>
    GF_up_avg{count} = GF_up/sign_Z;
    GF_dn_avg{count} = GF_dn/sign_Z;
    n_up_dn_avg{count} = n_up_dn/sign_Z;
    for m = 1:3
        two_point_corr_avg{count}{m} = two_point_corr{m}/sign_Z;
    end
    num_samples_avg(count) = num_samples;
    % average sign of fermion determinants, Z, where 
    % Z = 1/(Det(G_up)Det(G_dn)) for a given Hubbard-Stratonvich field
    sign_avg(count) = sign_Z/num_samples; 
end

% expectation values for each Markov chain is stored in a cell structure
% for final statistical average in a separate function
correlations_tmp = cell(1,4);
correlations_tmp{1} = GF_up_avg;
correlations_tmp{2} = GF_dn_avg;
correlations_tmp{3} = n_up_dn_avg;
correlations_tmp{4} = two_point_corr_avg;

% saving average spin and of number of samples related to partition
% function in another cell structure.
par_func_tmp = cell(1,2);
par_func_tmp{1} = num_samples_avg;
par_func_tmp{2} = sign_avg;

end
%%%%%%%%%%%%%%%%%%%%%%%%%%%%%%%%%%%%%%%%%%%%%%%%%%%%%%%%%%%%%%%%%%%%%%%%%%%%%%%%
\end{verbatim}
\end{tiny}

\begin{tiny}
\begin{verbatim}
%%%%%%%%%%%%%%%%%%%%%%%%%%%%%%%%%%%%%%%%%%%%%%%%%%%%%%%%%%%%%%%%%%%%%%%%%%%%%%%%
function [B,inv_B] = cluster(B_k,inv_B_k,h,n_l)

% Here we compute B_s(l) = exp(-Kdt)exp(-h_s(l)), h_up = -h_dn = h
% We also compute inv(B_s(l)) = exp(h_s(l))exp(Kdt) 
h = sparse(h);
B = cell(n_l,1);
inv_B = cell(n_l,1);
for l = 1:n_l
    h_tmp = h(l,:);
    B{l} = B_k*diag(exp(h_tmp));
    inv_B{l} = diag(exp(-h_tmp))*inv_B_k;
end

end
%%%%%%%%%%%%%%%%%%%%%%%%%%%%%%%%%%%%%%%%%%%%%%%%%%%%%%%%%%%%%%%%%%%%%%%%%%%%%%%%
\end{verbatim}
\end{tiny}

\begin{tiny}
\begin{verbatim}
%%%%%%%%%%%%%%%%%%%%%%%%%%%%%%%%%%%%%%%%%%%%%%%%%%%%%%%%%%%%%%%%%%%%%%%%%%%%%%%%
function [G_p,G_m,log_det_G,sign_det_G] = initial_eval(B_up_hat,B_dn_hat,k_vec,n_l)
                                        
% Here we use startification & QR decomposition
% to compute the GF from scratch

% spin up:
[G_p,log_det_G_p,sign_det_G_p] = strat(B_up_hat,k_vec,n_l);

% spin down:
[G_m,log_det_G_m,sign_det_G_m] = strat(B_dn_hat,k_vec,n_l);

% log of asb val of det(G_up*G_dn)
log_det_G = log_det_G_p + log_det_G_m; 
% sign of abs val of det(G_dn*G_dn)
sign_det_G = sign_det_G_p*sign_det_G_m;

end
%%%%%%%%%%%%%%%%%%%%%%%%%%%%%%%%%%%%%%%%%%%%%%%%%%%%%%%%%%%%%%%%%%%%%%%%%%%%%%%%
\end{verbatim}
\end{tiny}

\begin{tiny}
\begin{verbatim}
%%%%%%%%%%%%%%%%%%%%%%%%%%%%%%%%%%%%%%%%%%%%%%%%%%%%%%%%%%%%%%%%%%%%%%%%%%%%%%%%
function [G,log_det_G,sign_det_G] = strat(B,k_vec,n_l)

% computing Green's function via G = inv(Id + B_L...B_1) using QR
% decomposition for stabilization
% We also compute the sign and (log of ) abs value of GF's determinant

Q = 1;
D = 1;
T = 1;
i = 1;
while i <= n_l
    
    % stratification (i.e. multiplying dk B matrices directly
    % (no QR is needed as long as dk is small e.g. dk = 8 or 10))
    B0 = B{i};
    i = i +1;
    while i <= n_l && k_vec(i) < 1 
        B0 = B{i}*B0;
        i = i + 1;
    end
    c = (B0*Q)*D;

    % QR decomposition to multiply c to the accumulated 
    % product of c (and as a result B matrices) up to the current point 
    [Q,R] = qr(c);
    D0 = sparse(diag(R));
    inv_D = diag(1./D0);
    D = diag(D0);
    T = (inv_D*R)*T; 
    
end

% Using D, T, and Q we finally compute 
% GF with special care (to avoid numerical instability)
D_diag = diag(D);
D_b = max(1,abs(D_diag)).*sign(D_diag);
D_s = min(1,abs(D_diag)); 

inv_D_b = diag(sparse(1./D_b));
D_s = diag(sparse(D_s));

a1 = inv_D_b*(Q.') + D_s*T;
a2 = inv_D_b;
a3 = (Q.');

G = (a1\a2) * a3;


% computing log(det(abs(G))) as well as sign(det(G)) accurately 
% and with special care (again to avoid numerical instabilities)
[L1,U1] = lu(a1);

log_det_L1 = 0;
sign_det_L1 = sign(det(L1));
log_det_U1 = sum(log(abs(diag(U1))));
sign_det_U1 = prod(sign(diag(U1)));
log_det_a1 = log_det_L1 + log_det_U1;
sign_det_a1 = sign_det_L1*sign_det_U1;

log_det_a2 = -sum(log(abs(D_b)));
sign_det_a2 = prod(sign(D_b));

log_det_a3 = 0;
sign_det_a3 = sign(det(a3));

sign_det_G = sign_det_a1*sign_det_a2*sign_det_a3; % sign of det(G)
log_det_G = -log_det_a1+log_det_a2+log_det_a3; % log of abs(det(G))

end
%%%%%%%%%%%%%%%%%%%%%%%%%%%%%%%%%%%%%%%%%%%%%%%%%%%%%%%%%%%%%%%%%%%%%%%%%%%%%%%%
\end{verbatim}
\end{tiny}

\begin{tiny}
\begin{verbatim}
%%%%%%%%%%%%%%%%%%%%%%%%%%%%%%%%%%%%%%%%%%%%%%%%%%%%%%%%%%%%%%%%%%%%%%%%%%%%%%%%
function [G_up,G_dn,h,sgn_accumulated_tmp] = Sher_Mor(G_up,G_dn,h,n_s)

id = eye(n_s);
sgn_accumulated_tmp = 1;

for i = 1:n_s
    if h(1,i) ~= 0
        % ratio of 1/det(G_up)
        alpha_up = exp(-2*h(1,i))-1;
        r_up = 1+alpha_up*(1-G_up(i,i));

        % ratio of 1/det(G_dn)
        alpha_dn = exp(+2*h(1,i))-1;
        r_dn = 1+alpha_dn*(1-G_dn(i,i));

        % total ratio
        %=====================================
        r = r_up*r_dn;
        %=====================================

        % Hubbard-Stratonovic field flip acceptance based on Metropolis-Hastings algorithm
        % & as a consequence Green's function update based on the
        % Sherman-Morrison update formula
        if rand(1) <= abs(r)

            % spin up GF update:
            a_up = (id - G_up);
            b_up = G_up;
            G_up = G_up - (alpha_up/r_up)*a_up(:,i)*b_up(i,:);

            % spin dn GF update:
            a_dn = (id - G_dn);
            b_dn = G_dn;
            G_dn = G_dn - (alpha_dn/r_dn)*a_dn(:,i)*b_dn(i,:);

            % update h of accumulated sign of det
            h(1,i) = -h(1,i);
            sgn_accumulated_tmp = sgn_accumulated_tmp * sign(r);
        end
    end    
end

end
%%%%%%%%%%%%%%%%%%%%%%%%%%%%%%%%%%%%%%%%%%%%%%%%%%%%%%%%%%%%%%%%%%%%%%%%%%%%%%%%
\end{verbatim}
\end{tiny}

\begin{tiny}
\begin{verbatim}
%%%%%%%%%%%%%%%%%%%%%%%%%%%%%%%%%%%%%%%%%%%%%%%%%%%%%%%%%%%%%%%%%%%%%%%%%%%%%%%%
function [G_up,G_dn,h,sgn_accumulated_tmp] = delayed_update(G_up,G_dn,h,n_s,delayed_period)

% right update
id = eye(n_s);
sgn_accumulated_tmp = 1;

% right update
a_up = zeros(n_s,delayed_period);
b_up = zeros(delayed_period,n_s);
d_up = diag(G_up);
a_dn = a_up;
b_dn = b_up;
d_dn = diag(G_dn);

k = 1;
for i = 1:n_s
    if h(1,i) ~= 0
        % ratio of 1/det(G_up)
        alpha_up = exp(-2*h(1,i))-1;
        r_up = 1+alpha_up*(1-d_up(i));

        % ratio of 1/det(G_dn)
        alpha_dn = exp(2*h(1,i))-1;
        r_dn = 1+alpha_dn*(1-d_dn(i));

        % total ratio
        %============================================================
        r = r_up*r_dn;
        %============================================================

        % Hubbard-Stratonic field flip acceptance based on Metropolis-Hasting algorithm
        % and as a consequence Green's function update based on the
        % Sherman-Morrison-Woodbury update formula
        if rand(1) <= abs(r)
            v1 = 1:(k-1);

            % spin up GF update:
            a_up(:,k) = G_up(:,i)-id(:,i) + a_up(:,v1)*b_up(v1,i);
            a_up(:,k) = (alpha_up/r_up)*a_up(:,k);
            b_up(k,:) = G_up(i,:) + a_up(i,v1)*b_up(v1,:);
            d_up = d_up + a_up(:,k).*b_up(k,:).';

            % spin dn GF update:
            a_dn(:,k) = G_dn(:,i) -id(:,i)+ a_dn(:,v1)*b_dn(v1,i);
            a_dn(:,k) = (alpha_dn/r_dn)*a_dn(:,k);
            b_dn(k,:) = G_dn(i,:) + a_dn(i,v1)*b_dn(v1,:);
            d_dn = d_dn + a_dn(:,k).*b_dn(k,:).';

            k = k+1;
            h(1,i) = -h(1,i);

            sgn_accumulated_tmp = sgn_accumulated_tmp * sign(r);
            
        end
        if k == delayed_period+1
            G_up = G_up + a_up*b_up;
            d_up = diag(G_up);

            k = 1;

            G_dn = G_dn + a_dn*b_dn;
            d_dn = diag(G_dn);

        end
    end
end

v1 = 1:(k-1);
G_up = G_up + a_up(:,v1)*b_up(v1,:);
G_dn = G_dn + a_dn(:,v1)*b_dn(v1,:);

end
%%%%%%%%%%%%%%%%%%%%%%%%%%%%%%%%%%%%%%%%%%%%%%%%%%%%%%%%%%%%%%%%%%%%%%%%%%%%%%%%
\end{verbatim}
\end{tiny}

\begin{tiny}
\begin{verbatim}
%%%%%%%%%%%%%%%%%%%%%%%%%%%%%%%%%%%%%%%%%%%%%%%%%%%%%%%%%%%%%%%%%%%%%%%%%%%%%%%%
function [correlations,measurements] = exp_val(par_func_tmp,...
      correlations_tmp,n_markov,U,mu,n_x,n_y,t_x,t_y,t_xy,bnd_x,bnd_y)

n_s = n_x*n_y; % number of sites
hopping = hopping_matrix(t_x,t_y,t_xy,n_x,n_y,bnd_x,bnd_y);% hopping matrix

% combining results of all Markov chains
GF_up_avg = correlations_tmp{1};
GF_dn_avg = correlations_tmp{2};
n_up_dn_avg = correlations_tmp{3};
two_point_corr_avg = correlations_tmp{4};

sign_avg = par_func_tmp{2};

energy_avg = zeros(n_markov,1);
for count = 1:n_markov
    % average ground-state E
    energy_avg(count,1) = (U*sum(n_up_dn_avg{count})+...
        trace(hopping*(GF_up_avg{count}+GF_dn_avg{count})))/n_s;    
end

energy_mean = 0;
sign_mean = 0;
GF_up_mean = 0;
GF_dn_mean = 0;
n_up_dn_mean = 0;
two_point_corr_mean = cell(1,3);
for l = 1:3
    two_point_corr_mean{l} = 0;
end

for count = 1:n_markov
    energy_mean = energy_mean + energy_avg(count,1);
    sign_mean = sign_mean + abs(sign_avg(count,1));
    GF_up_mean = GF_up_mean + GF_up_avg{count};
    GF_dn_mean = GF_dn_mean + GF_dn_avg{count};
    n_up_dn_mean = n_up_dn_mean + n_up_dn_avg{count};
    for l = 1:3
        two_point_corr_mean{l}=two_point_corr_mean{l}+...
            two_point_corr_avg{count}{l};
    end
end

energy_mean = energy_mean/n_markov;
sign_mean = sign_mean/n_markov;
GF_up_mean = GF_up_mean/n_markov;
GF_dn_mean = GF_dn_mean/n_markov;

n_up_mean = diag(GF_up_mean);
n_dn_mean = diag(GF_dn_mean);
n_up_dn_mean = n_up_dn_mean/n_markov;
for l = 1:3
    two_point_corr_mean{l} = two_point_corr_mean{l}/n_markov;
end

% saving results in a cell structure
correlations = cell(1,10);
correlations{1}{1} = sign_avg;
correlations{2}{1} = GF_up_mean;
correlations{2}{2} = GF_dn_mean;
correlations{3} = n_up_dn_mean;
correlations{4} = two_point_corr_mean;
correlations{5} = energy_avg;

% average density
n_mean = mean(n_up_mean)+mean(n_dn_mean);

% average kinetic energy
kin_energy = energy_mean -U*mean(n_up_dn_mean);

% average Hubbard interaction energy
int_energy = U*mean(n_up_dn_mean);

% normalized statistical error bar of DQMC calculations
err_bar = 10^2*std(energy_avg)/abs(mean(energy_avg));

% most important measurements

fprintf('measurements=[density,energy,kin_energy,int_energy,err_bar,<sign>]');

measurements=[n_mean,energy_mean,kin_energy,int_energy,err_bar,sign_mean]

end
%%%%%%%%%%%%%%%%%%%%%%%%%%%%%%%%%%%%%%%%%%%%%%%%%%%%%%%%%%%%%%%%%%%%%%%%%%%%%%%%
\end{verbatim}
\end{tiny}

}

\end{widetext}
\end{document}